\documentclass[fleqn,usenatbib]{mnras}

\usepackage{newtxtext,newtxmath}

\usepackage[T1]{fontenc}

\DeclareRobustCommand{\VAN}[3]{#2}
\let\VANthebibliography\thebibliography
\def\thebibliography{\DeclareRobustCommand{\VAN}[3]{##3}\VANthebibliography}

\usepackage[T1]{fontenc}
\usepackage{ae,aecompl}
\usepackage{longtable}
\usepackage{lipsum}
\usepackage{lscape}
\usepackage{amsmath}

\usepackage{graphicx}	
\usepackage{amsmath}	
\usepackage{float}
\usepackage{subfig}
\bibpunct{(}{)}{;}{a}{}{,} 
\usepackage{float}
 \usepackage{longtable}
 
\newcommand{\xmm}{{\it XMM-Newton}}
\newcommand{\chandra}{{\it Chandra}}

\title[\xmm\, study of the Sculptor dSph]{\xmm\, study of the Sculptor dwarf spheroidal galaxy}

\author[Sara Saeedi et al]{
Sara Saeedi,$^{1}$\thanks{E-mail: sara.saeedi@fau.de}
Manami Sasaki,$^{1}$
\\
$^{1}$ Dr. Karl Remeis-Sternwarte, Erlangen Centre for Astroparticle Physics, Friedrich-Alexander-Universit\"at Erlangen-N\"urnberg,\\ Sternwartstrasse 7, 96049, Bamberg, Germany\\
}

\date{Accepted XXX. Received YYY; in original form ZZZ}

\pubyear{2015}

\begin{document}
\label{firstpage}
\pagerange{\pageref{firstpage}--\pageref{lastpage}}
\maketitle

\begin{abstract}
We analysed  observations with \xmm\, in the field of the Sculptor dwarf spheroidal galaxy (dSph). The aim of the study was the classification of X-ray binaries and accreting white dwarfs, which belong to the Sculptor dSph. Using different methods of X-ray timing and spectral analyses, together with an extensive multi-wavelength study of the optical and infrared counterparts of the X-ray sources, we classified the sources detected with \xmm\ in the field of Sculptor dSph.  
The long term variability of the sources has been studied over  two \xmm\, observations. None of the members of Sculptor dSph  show a significant long-term variability over these two observations.
We also searched for periodicity and pulsation using the Lomb-Scargle and Rayleigh's Z$^{2}_{n}$ techniques. No signal of pulsation or periodicity have been found for the X-ray sources. The results show the presence of a noticeable number of background X-ray sources in the field of this galaxy. We classified  43 sources as active galactic nuclei (AGNs), galaxies, and galaxy candidates. Three Galactic foreground stars have been identified in the field of Sculptor dSph, and one of them is an M-dwarf candidate. Moreover, we classified  four symbiotic star candidates and three quiescent low-mass X-ray binary candidates in Sculptor dSph. The luminosity of these X-ray sources is $\sim10^{33-35}$~erg\,s$^{-1}$. 
\end{abstract}




\section{Introduction}
\label{intro}
X-ray observations play a key role in studying the evolution of different types of galaxies. They unveil the properties of populations of discrete X-ray sources, which are mostly X-ray binaries (XRBs) or accreting white dwarfs (AWDs). Most of the satellite galaxies of the Milky Way are dwarf spheroidal galaxies  \citep[dSphs;][]{2013pss5.book.1039W}. As many dSphs show no recent star formation, they are ideal targets to study the old stellar populations in galaxies formed in the early stages of galaxy evolution. Considering the close distances of many dSphs, they are ideal targets to detect and identify low luminosity X-ray sources, e.g., AWDs or quiescent low-mass X-ray binaries (QLMXBs) \citep[10$^{33}-10^{35}$~erg\,s$^{-1}$;][]{2017PASP..129f2001M}. As a part of an observational project we started to study the population of dwarf galaxies in X-rays to determine the nature of their low-luminosity X-ray sources and also to study their source population considering differences in the mass, age, and the metallicity of these galaxies.

 So far, the X-ray population of some dSphs has been studied using the data of \xmm\, and \chandra\, observations. Among the dSphs, Fornax dSph has the most recent star formation history. Fornax dSph experienced multiple epochs of star formation \citep{2008ApJ...685..933C} and has a dominant population of intermediate-age stars. Studies show that the last star formation events occurred 1 Gyr ago and continued until 100--200 Myr ago \citep[][]{2012A&A...544A..73D, 2013MNRAS.433.1505D}. \xmm\, observation has been used for the source classification of Fornax \citep{2010AIPC.1314..337O, 2013A&A...550A..18N}.  A low-mass X-ray binary (LMXB) candidate was found about 0.6 arcmin from the center of the globular cluster CG 4 in Fornax dSph \citep{2010AIPC.1314..337O}. Two X-ray sources are coincident with the two Fornax globular clusters GC 3 and GC\,4, although their nature is not known \citep{2013A&A...550A..18N}. Both works focused on the classification of LMXBs in this galaxy, however, no systematic classification of all X-ray sources was presented. Leo I dSph had a continuous star formation from 15 Gyr to 0.5~Gyr ago and its largest epoch of star formation occurred recently, from 3 to 1~Gyr ago \citep{2009A&A...500..735G,2010MNRAS.404.1475H}. \citet{2010AIPC.1314..337O} suggested some candidates for LMXBs in Leo~I dSph, which had  yet not been confirmed. Moreover, with the aim of searching for high-mass X-ray binaries \citep[related to the recent star formation,][]{2005astro.ph..6430D}, the X-ray population of the galaxy was studied \citep{2012MNRAS.422.2302B}. No high-mass X-ray binary was classified in the galaxy. The star formation in  Carina dSph is believed to have happened $\sim$10 to $\sim$6 Gyr ago \citep{1994AJ....108..507S}.  Also, Sagittarius dSph, as one of the most massive dSphs \citep{2012AJ....144....4M} is dominated by an intermediate-age population \citep[between 6 and 9~Gyr;][]{2006A&A...446L...1B}. Carina  dSph and the center of Sagittarius dSph have been studied by \citet{2006A&A...459..777R}. The study did not confirm any specific X-ray source as a member of these two dSphs, while it was reported
that there should be fraction of unclassified soft X-ray sources in Carina and Sagittarius dSphs \citep{2006A&A...459..777R}. Ursa Major II \citep{2012ApJ...752...42D}, Draco \citep{2007MNRAS.375..831S}, Willman~1 \citep{2005AJ....129.2692W, 2011AJ....142..128W}, Ursa Minor  \citep{2002AJ....123.3199C}, and Sculptor \citep[e.g,][]{1999PASP..111.1392M} have a  population, which has mainly been formed $>10$~Gyr ago. Among these galaxies, the X-ray population of  Draco, Leo I, Ursa Major II, and Ursa Minor has been studied by \citet{2015MNRAS.451.2735M}  searching for an intermediate mass black hole. 
No candidate has been detected in these dSphs.  \citet{2020MNRAS.499.3111S} confirmed the presence of a symbiotic star in Willman~1 and also suggested possible candidates for QLMXBs in this dSph. 
 Among the dSphs, Draco had the deepest X-ray observation with \xmm. Two classification studies have been performed for the Draco dSph using the \xmm\, observations of 2009 \citep{2016A&A...586A..64S, 2016ApJ...821...54S}. \citet{2016ApJ...821...54S} suggested some candidates for the LMXBs and AWDs in the Draco dSph. In the study of the X-ray population of the Draco dSph, \citet{2016A&A...586A..64S} provided information about the counterparts of the X-ray sources using  multi-wavelength data and could clearly separate the background sources (AGNs and galaxies) from the sources with stellar counterparts \citep{2016A&A...586A..64S}. The incompleteness-corrected X-ray luminosity function (XLF) of X-ray sources shows that the main sources in the Draco dSph are soft X-ray sources with low luminosities, e.g, AWDs \citep{2016A&A...586A..64S}. Using deep \xmm\, observations of 2015 we focused on the classification of AWDs, which are expected to be the most numerous X-ray sources in the old population of Draco dSph.  We classified and studied four AWDs of the Draco dSph \citep{2019A&A...627A.128S}. 
We decided to extend our study to the Sculptor dSph (RA=$01^{\mathrm h} 00^{\mathrm m} 09.3^{\mathrm s}$, Dec=$-33^{\circ} 42\arcmin 33\arcsec$ [J2000]) at a distance of $\sim$86 kpc and a stellar mass of $2.3\times10^{6}~M_{\sun}$, which makes it the dSph with the highest stellar mass within the distance $<100$~kpc \citep{2012AJ....144....4M}. The optical studies confirm a very old stellar population for Sculptor dSph, similar to that of  Draco dSph \citep[][]{2014ApJ...789..147W, 2019MNRAS.487.5862B}.  Sculptor dSph has been observed with \chandra\, using the ACIS-S camera (size of the field of view = $8\farcm3\times33\farcs2$) with 21 exposures of 6 ks each. The aim of the \chandra\, observations was to detect bright XRBs in this galaxy and five candidates were reported \citep{2005MNRAS.364L..61M}. \citet{2019MNRAS.485.2259A} performed a multi-wavelength study using Spitzer and Gemini telescopes for thirteen bright X-ray sources in the field of Sculptor dSph including those five  sources, which have been suggested as candidates for low-mass X-ray binaries (LMXBs) in the study of \citet{2005MNRAS.364L..61M}. The study of \citet{2019MNRAS.485.2259A} did not confirm any source as a member of Sculptor dSph. 

In this work, we have performed a multi-wavelength study of X-ray sources detected in two \xmm\, observations, which cover a larger region and also have  a higher exposure  time in comparison to those of \chandra. These allowed us to perform X-ray timing and spectral analysis for the bright X-ray sources in the field of Sculptor dSph. Moreover, we  re-consider the X-ray spectral analysis of those sources, which have been discussed in the study of \citet{2019MNRAS.485.2259A}.

In following sections, first we explain the process of \xmm\, data reduction and catalogue preparation (Sect.~\ref{data-red}). Afterwards, the details of the X-ray analysis (Sect.\ref{data-ana}) and multi-wavelength (optical/infrared) studies (Sect.~\ref{multi}) are explained. In Section~\ref{diss}, we discuss the details of source classification.

\section{\xmm\, data of Sculptor}
\label{data-red}
\subsection{Data reduction}
Sculptor has been observed twice with \xmm\ in 2019. One can see the details of the observation in Table\,\ref{obs-data}. The \xmm\ cameras EPIC-pn \citep{2001A&A...365L..18S} and EPIC-MOS1,\,2 \citep{2001A&A...365L..27T} were in full-frame mode. The medium filter was used during the observations. We have performed  data reduction and source detection using the \xmm\, Science Analysis System\,(\texttt{SAS},\,V.20.0.0). In the first step, the event files of observations were filtered to remove intervals with a high background caused by soft proton flares. We applied a threshold rate of $\leq$ 0.35 count\,s$^{-1}$ for EPIC-MOS and  $\leq$ 0.4 count\,s$^{-1}$ for EPIC-pn to define the good time intervals.  The net exposure times for each observation and EPIC camera are listed in  Table\,\ref{obs-data}. Source detection was performed using the \texttt{SAS}
task \texttt{edetect-chain} for each observation in the five standard energy-bands of \xmm\, B1\,(0.2--0.5\,keV), B2\,(0.5--1.0\,keV), B3\,(1.0--2.0\,keV), B4\,(2.0--4.5\,keV), B5\,(4.5--12.0\,keV) . We applied a minimum value for the maximum likelihood\,($L$) of 10 for the source detection. The detection maximum likelihood $L$=--ln$(p)$ is defined as the probability of Poisson random fluctuations of the counts\,($p$), which is based on the raw counts of the source versus the raw counts of the local background.
\begin{table}
    \caption{\xmm\, observations of Sculptor dSph. \label{obs-data}}
    \small
     \begin{tabular}{cccccc}
\hline\hline
OBS-N0 & OBS-ID & OBS-Date &  \multicolumn{3}{c}{EXP.T${^\ast}$ (ks)} \\
       &        &           &  pn&MOS1&MOS2        \\
\hline
       1& 0844430101   & 2019-06-03   &  38.2  &    41.7 & 41.7      \\
       2&  0844430201 &  2019-11-27  & 37.6   & 41.7    & 41.6       \\
      \hline
      \multicolumn{6}{l}{$\ast$: Exposure time of EPICs after screening for high background.}\\
     \end{tabular}
\end{table}
\subsection{Source Catalogue}
\begin{figure*}
\includegraphics[trim={0.cm 1.6cm 1.0cm 0.0cm},clip, width=0.80\textwidth]{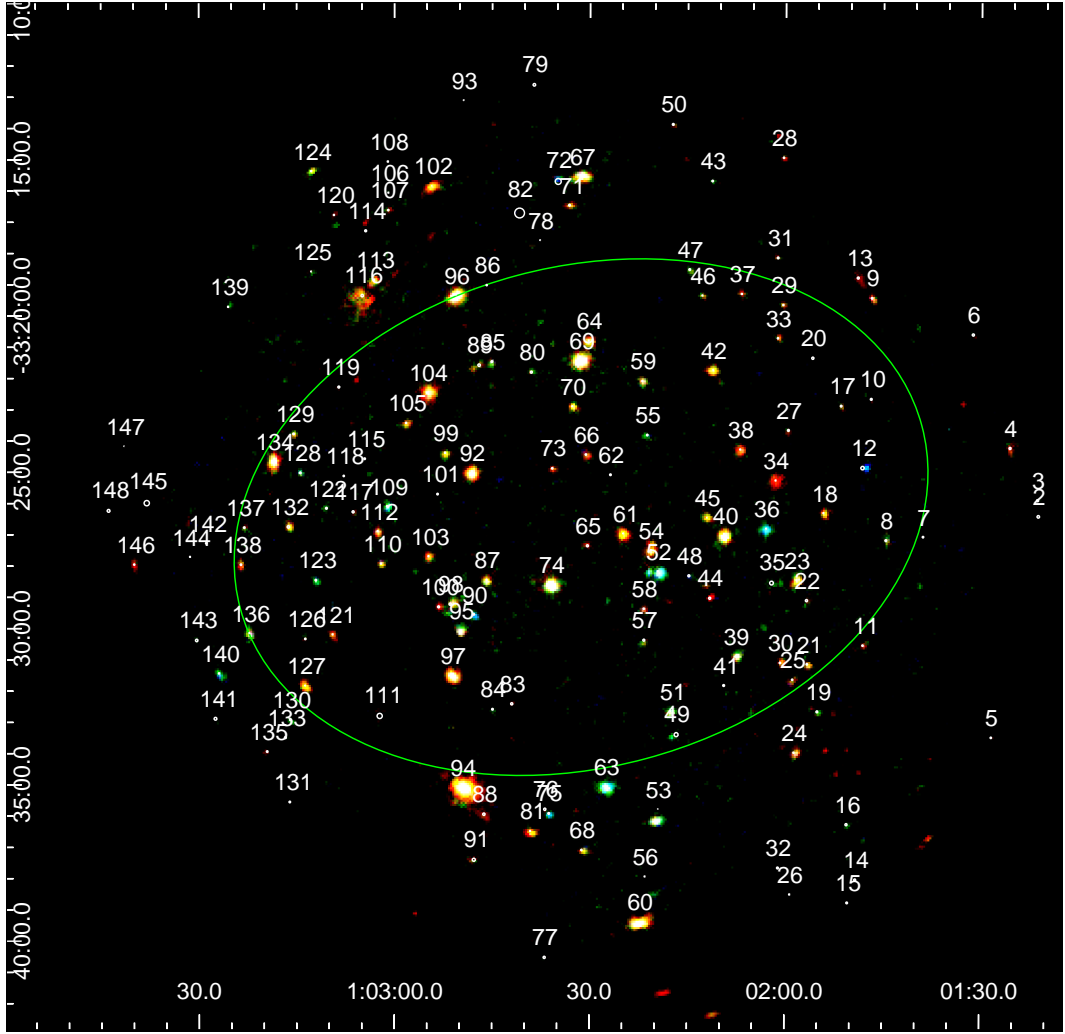}
\caption{ Three-colour mosaic image of Sculptor dSph. The position of the  detected sources are shown on the image. The green ellipse shows the main area of Sculptor dSph using the half-mass radius \citep{2012AJ....144....4M}. \label{rgb-image}}
\end{figure*}
The final source catalogue is obtained by cross-checking the sources detected in two observations and all EPICs. If sources, which were detected in both observations, were closer to each other than the 3$\sigma$ statistical positional errors, they are considered as one source. Artificial detections caused by bad pixels, hot columns, gaps, and edges of the CCD chips \citep{2008A&A...480..611S} were not considered as real sources.
These can be identified since they are detected only in one camera. The final catalogue of 148 X-ray sources in the field of Sculptor dSph is presented in Table~\ref{catalogue-x-ray}. The catalogue shows the source ID, right ascension\,(RA), declination\,(Dec), 1$\sigma$ positional uncertainty, the flux of different observations for each source, and the significant hardness ratios\,(HRs) of each source as explained in Sect.\ref{hr-sec}. The list of the sources is sorted by their coordinates. The ID of the sources in Table~\ref{catalogue-x-ray} is used to present the sources in this work.\\
\begin{table}
    \caption{Offsets of the \xmm\, observations \label{offset}}
     \begin{tabular}{clrr}
\hline\hline
OBS-N0 &  EPIC & $\Delta$RA ($\arcsec$) & $\Delta$DEC ($\arcsec$) \\
\hline
1  &PN       &  -0.89$\pm$0.12  &   -0.02$\pm$0.14 \\
   &MOS1     &  -0.33$\pm$0.56  &  -0.55$\pm$0.33 \\
   &MOS2     &  -0.70$\pm$0.52  &   -0.82$\pm$0.32 \\
2  &PN       &  0.02$\pm$0.10  &   0.21$\pm$0.15 \\
   &MOS1     & -0.23$\pm$0.41  &  -0.11$\pm$0.64 \\
   &MOS2     &  0.14$\pm$0.33  &   -0.53$\pm$0.38 \\
      \hline
     \end{tabular}
\end{table}
The mosaic images were created using the \texttt{SAS} task \texttt{emosaicproc} using the calibrated event files of both observations and all EPICs. Using this task the event files of different observations with different pointings are centered into one image and source detection routine of \texttt{SAS} was applied to detect the sources in the mosaic image.  Figure\,\ref{rgb-image} shows the three-colour mosaic image of all observations. The position of the detected sources and the half mass radius\,($r_{\rm h}$) of Sculptor dSph \citep{2012AJ....144....4M} are shown in Fig~\ref{rgb-image}. 

For the astrometric correction of the position of X-ray sources, we selected 25 sources, which have been already classified as AGNs in optical catalogues (see Sect.\ref{AGN-cata}). We calculated the weighted mean differences of $\Delta$RA and  $\Delta$Dec between the positions of the optical and X-ray sources to find an offset for the position of X-ray sources for each observation (see Table~\ref{offset}). If the weighted mean positional difference was significant the offset has been applied to correct the coordinate of sources for each camera. After the offset, the coordinate and the positional error of each source were taken from the observation, in which the source was detected with the highest maximum likelihood and reported in the final catalogue (Table.~\ref{catalogue-x-ray}).\\
As the HR study shows (see Fig.~\ref{hrs-plot}), the majority of X-ray sources in the field of Sculptor are mainly background sources with a power law spectrum with a photon index of $\Gamma \sim$2  and a very low absorption. To obtain the flux of the sources in each observation, we assumed an absorbed power law model with the Galactic foreground absorption in the direction of Sculptor dSph \citep[$N_{\rm H}$=1.75$\times10^{20}$~cm$^{-2}$,][]{2016A&A...594A.116H} and a photon index of $\Gamma$=2. The flux of the sources is calculated using the relation $F_{x}=\frac{RATE_{x}}{ECF}$, where the energy conversation factor\,(ECF) is calculated for each camera based on the above model. Table~\ref{ecf-tab} lists the ECFs of EPIC-pn and EPIC-MOS for each energy band. Table~\ref{catalogue-x-ray} presents the weighted mean flux of all EPICs for each observation. 

\begin{table}
    \caption{ECF for the EPIC-pn and EPIC-MOS cameras}
    \label{ecf-tab}
     \begin{tabular}{crr}
\hline\hline
  Energy bands & EPIC-pn & EPIC-MOS \\
  keV&          cts\,erg$^{-1}$\,cm$^{-2}$         &     cts\,erg$^{-1}$\,cm$^{-2}$        \\
\hline
0.2--0.5 &1.10e+12  & 1.60e+11\\
0.5--1.0  &8.25e+11  & 1.81e+11\\
1.0--2.0 &4.91e+11  & 1.95e+11\\
2.0--4.5 &0.78e+11  & 7.21e+10\\
4.5--12.0  &5.80e+10 & 1.78e+10\\
      \hline
     \end{tabular}
\end{table}
\section{X-ray data analysis}
\label{data-ana}
\subsection{X-ray timing analysis}
\label{X-time}
\begin{figure}
  \centering
\includegraphics[clip, trim={2.2cm 0.5cm 0.0cm 0.cm},width=0.48\textwidth]{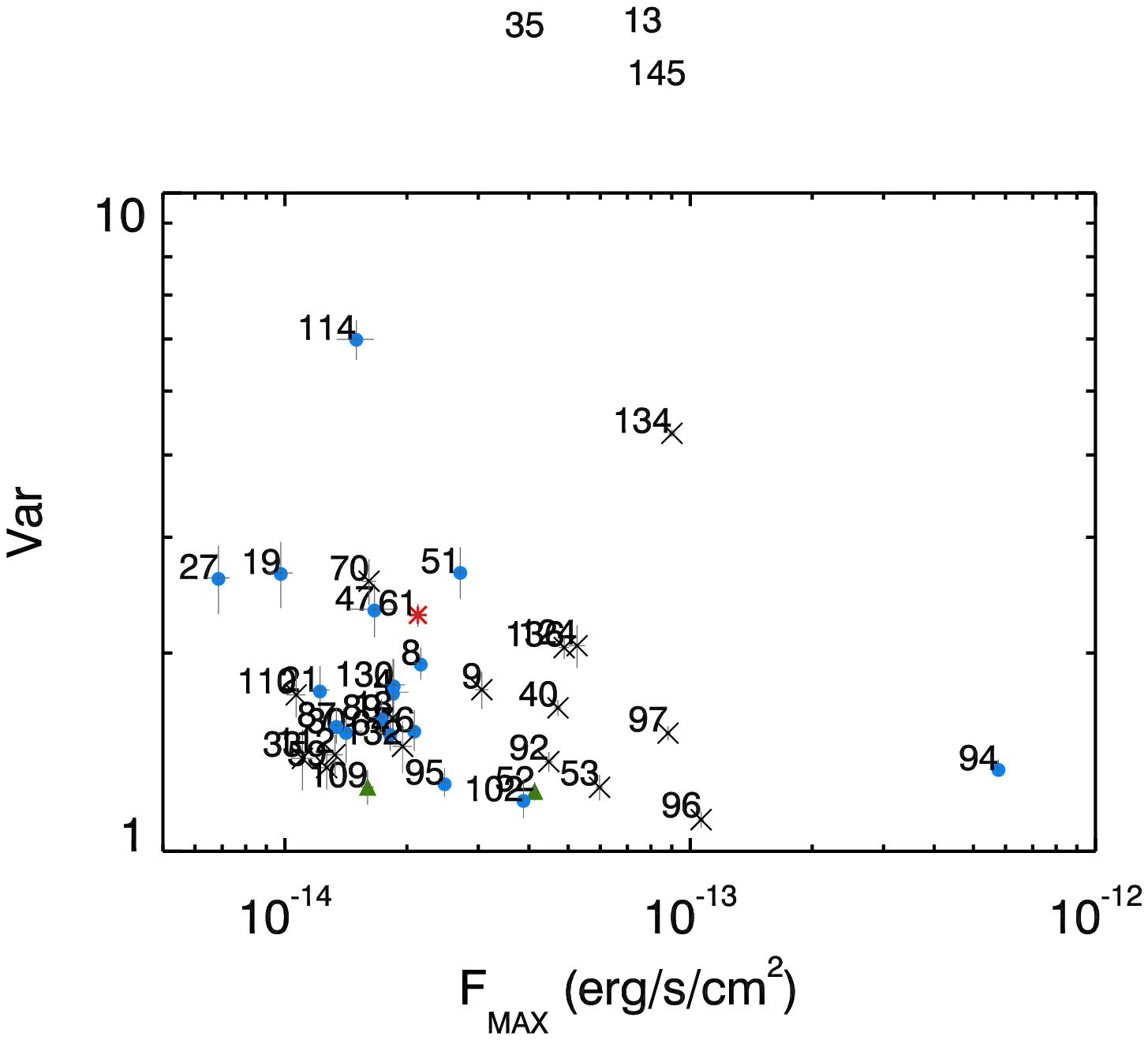}
\hspace{1cm}
\caption{Variability factor of sources with significant variability (S>3) in the energy band of 0.2–4.5 keV plotted versus the maximum flux.  In all plots, classified background AGNs are marked with black crosses,  classified background galaxies with blue squares (not included here), foreground stars with red asterisks, symbiotic stars with green triangles, and unclassified sources with blue circles. \label{var}}
\end{figure}

The X-ray timing analysis was based on two different studies of short-term variability   (pulsation) and long-term variability. For all unknown sources, which have been confirmed neither as foreground stars nor as AGNs in available catalogues, we searched for pulsation signals using the  Z$^{2}_{n}$  test \citep{1983A&A...128..245B, 1988A&A...201..194B}. For the unknown sources with  counts\,>\,300 in each observation, we created light curves and applied the  Lomb-Scargle technique \citep{1982ApJ...263..835S} to find a possible periodicity.  For the sources, which had a Sculptor member as a counterpart, we show the Lomb-Scargle periodograms in the Fig.~\ref{lomb}. The plots show the observed power for selected frequencies. We used the false alarm probability of 0.001 to detect signals with >3$\sigma$ confidence level \citep[see equation~18;][]{1982ApJ...263..835S}.  As the plots show none of the sources present a significant  amplitude as a signal of periodicity.\\
To study the long term variability, we checked the flux variation of sources over  two \xmm\, observations. Flux variation and its significance were calculated using 
\begin{equation} 
Var=\frac{F_{\rm max}}{F_{\rm min}}~~~\mathrm{and}~~~S=\frac{F_{\rm max}-F_{\rm min}}{\sqrt{EF_{\rm max}^{2}+EF_{\rm min}^{2} }}, 
\end{equation}
respectively {\citep{1993ApJ...410..615P}}. Here,  $F_{\rm max}$ and $F_{\rm min}$ are the maximum and minimum X-ray flux,  and $EF_{\rm min}$  and $EF_{\rm max}$ are their corresponding errors. We excluded the last \xmm\ energy band (4.5--12.0 keV) from the variability study due the high background fluctuations. For the sources, which have been detected in both observations, the variability factor was calculated (see Table~\ref{catalogue-x-ray}). Sources, for which $S$ was higher than 3, are considered as sources with significant variability. Figure~\ref{var} shows the sources with significant variability versus their maximum flux. Only an unknown source (src No.\,114) and an AGN (src No.\,134) show noticeable variability. Src No.\,114 has only a faint optical counterpart. The lack of both the X-ray spectrum and multi-wavelength counterparts for the source does not allow us to constrain its nature. None of the sources, which are candidates to be Sculptor members (see Sect.~\ref{can-diss}) show significant long-term variability. 

\subsection {Hardness ratios}
\label{hr-sec}
\begin{figure*}
  \centering
\includegraphics[clip, trim={0.2cm  0.5cm  0.5cm  0.cm},width=0.33\textwidth]{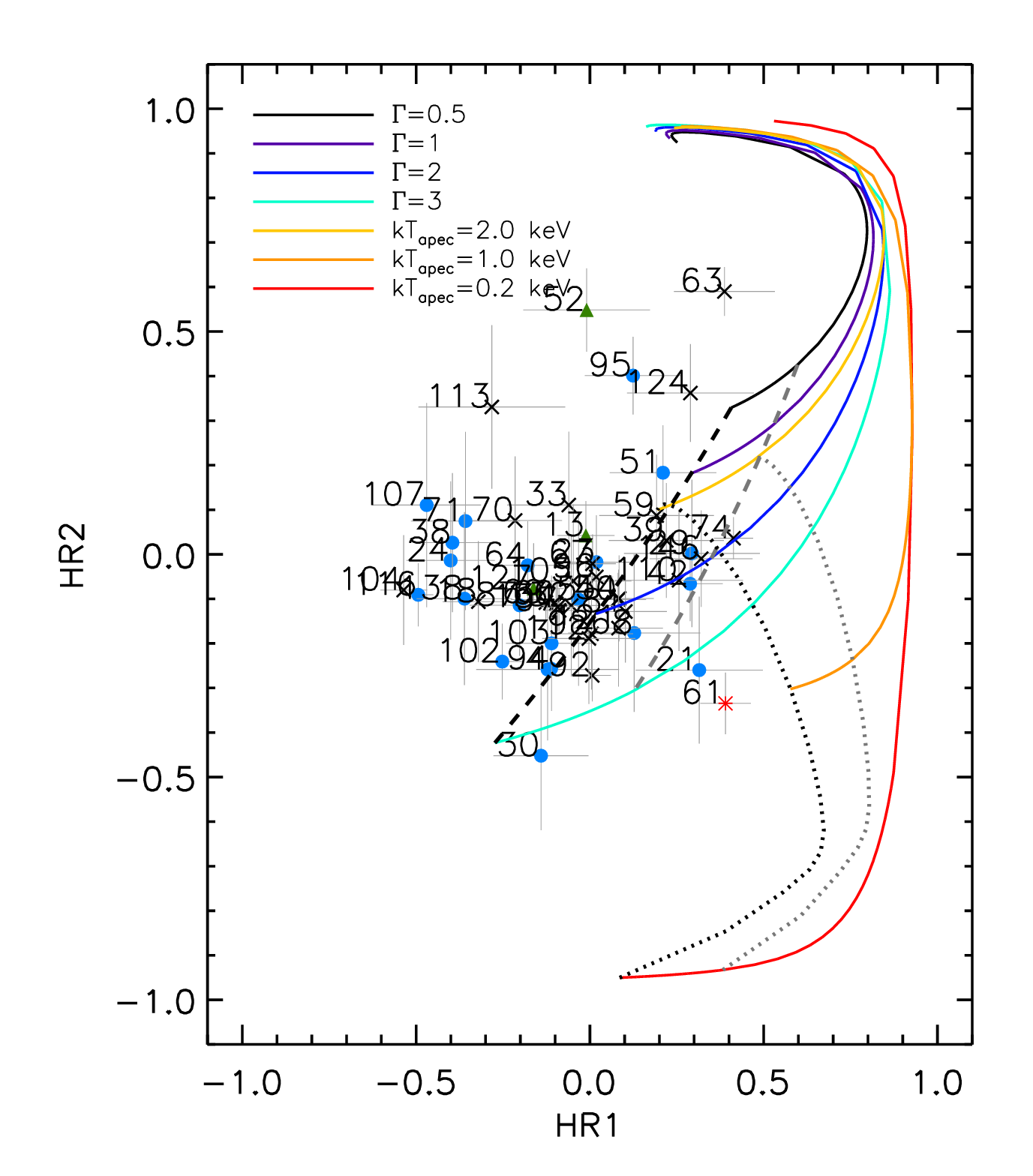}
\includegraphics[clip, trim={0.2cm  0.5cm  0.5cm  0.cm},width=0.33\textwidth]{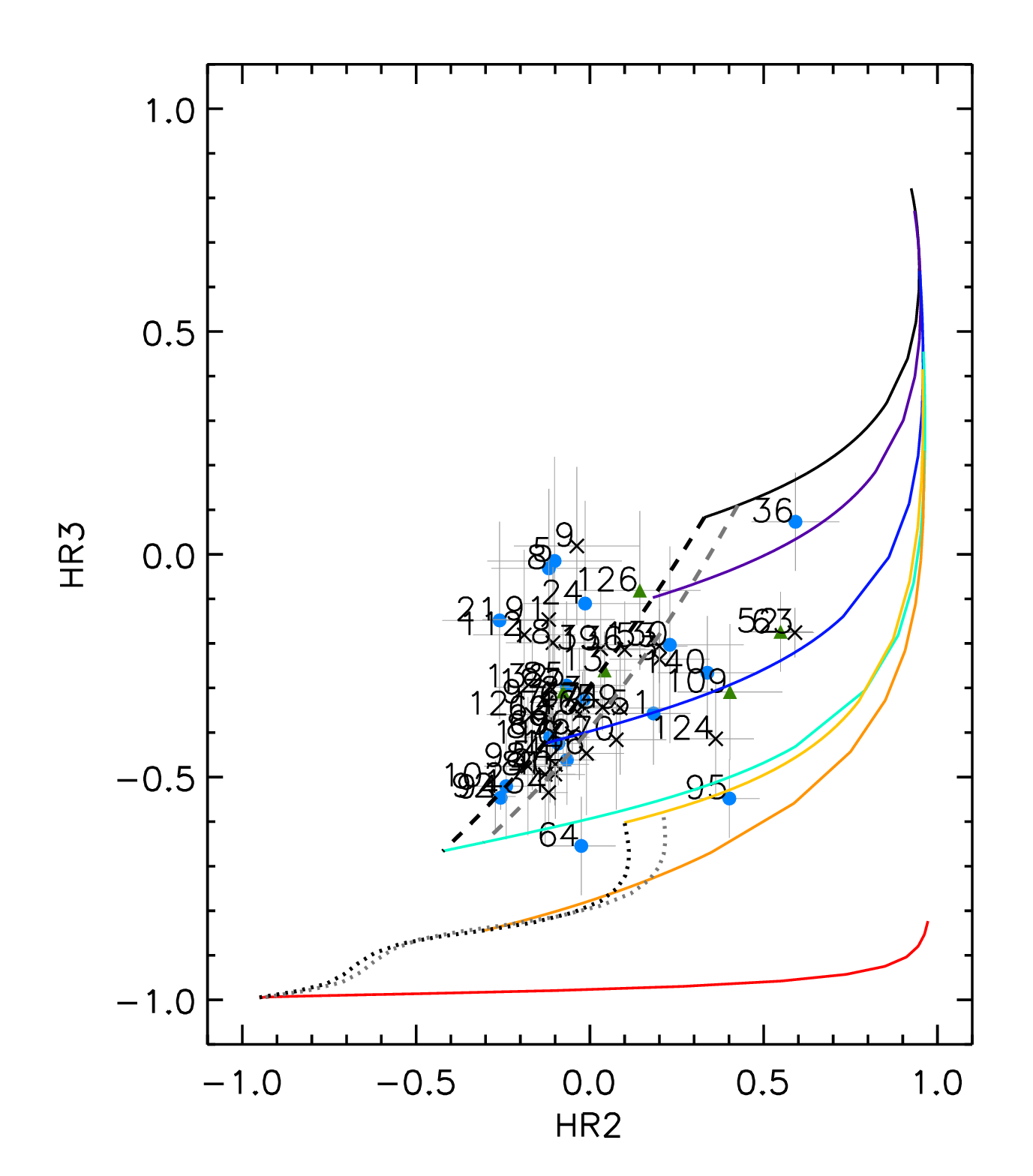}
\includegraphics[clip, trim={0.2cm  0.5cm  0.5cm  0.cm},width=0.33\textwidth]{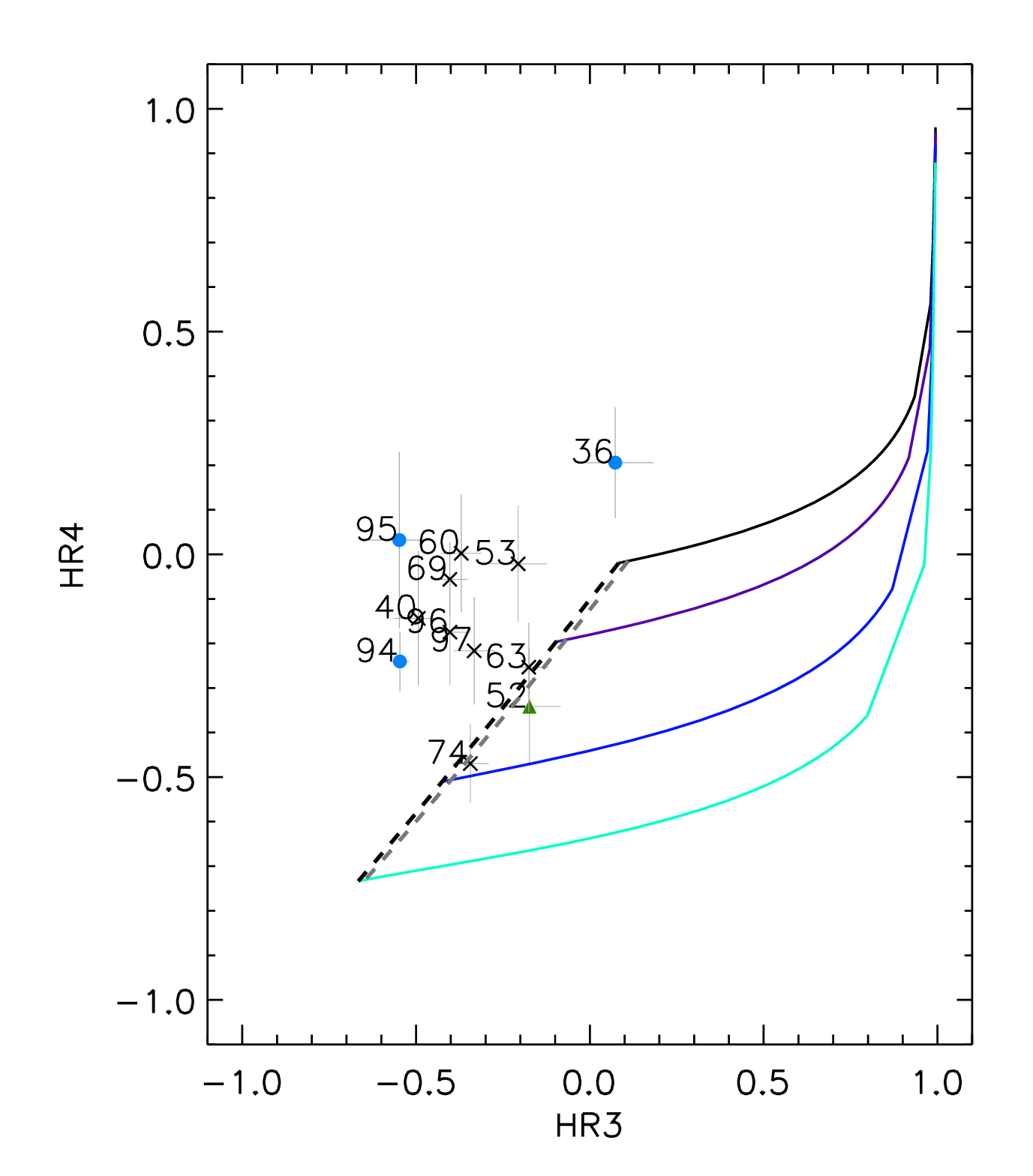}
\hspace{1cm}
\caption{Hardness ratio diagrams. The plotted solid lines are the hardness ratios calculated for different spectral models and column densities. For the power-law models (dashed lines) and apec models (dotted lines), the  column densities of $N_{\rm H}$=$10^{20}$\,cm$^{-2}$ (in black) and $N_{\rm H}$=$10^{21}$\,cm$^{-2}$ (in gray) are plotted.  The symbols are the same as in Fig.~\ref{var}. \label{hrs-plot}}
\end{figure*}

Hardness ratios are useful parameters for the study of the spectral properties of X-ray sources especially in case of faint sources, for which no X-ray spectrum can be extracted. The hardness ratio and its error are defined as 
\begin{equation} 
 HR_\mathrm{i}=\frac{B_\mathrm{i+1}-B_\mathrm{i}}{B_\mathrm{i+1}+B_\mathrm{i}} ~~~\mathrm{and}~~EHR_i=2\frac{\sqrt{(B_\mathrm{i+1}EB_\mathrm{i})^2+(B_\mathrm{i}EB_\mathrm{i+1})^2}} {(B_\mathrm{i+1}+B_\mathrm{i})^2}, 
\end{equation}
respectively, where $B_\mathrm{i}$ is the count rate and $EB_\mathrm{i}$ is the corresponding error in the band $i$.  We calculated the hardness ratio from the observation, in which the source had the highest detection likelihood.  To increase the accuracy of the measurement, we calculated the hardness ratio only for those energy bands, which had a detection likelihood higher than 6 (>3$\sigma$). Table~\ref{catalogue-x-ray} lists the HRs for the sources. Figure~\ref{hrs-plot} shows the HR diagrams. To compare with different spectral models, we plotted the lines representing the hardness ratios of different spectral models with various column densities from $N_{\rm H}$=$10^{20}$\,cm$^{-2}$ to $N_{\rm H}$=$10^{23}$\,cm$^{-2}$. Three \texttt{power-law} models with photon-index $\Gamma$ of 1, 2, 3 correspond to  hardness ratio of  hard sources, e.g, X-ray binaries or AGNs, and three \texttt{apec} model with the temperature of $kT$ of  0.2, 1.0, and 2.0 keV represent the spectra of soft plasma emissions detected in different sources, e.g, supernova remnant\,(SNR), foreground stars, and symbiotic stars.  
\subsection{Spectral analysis}
\begin{figure*}
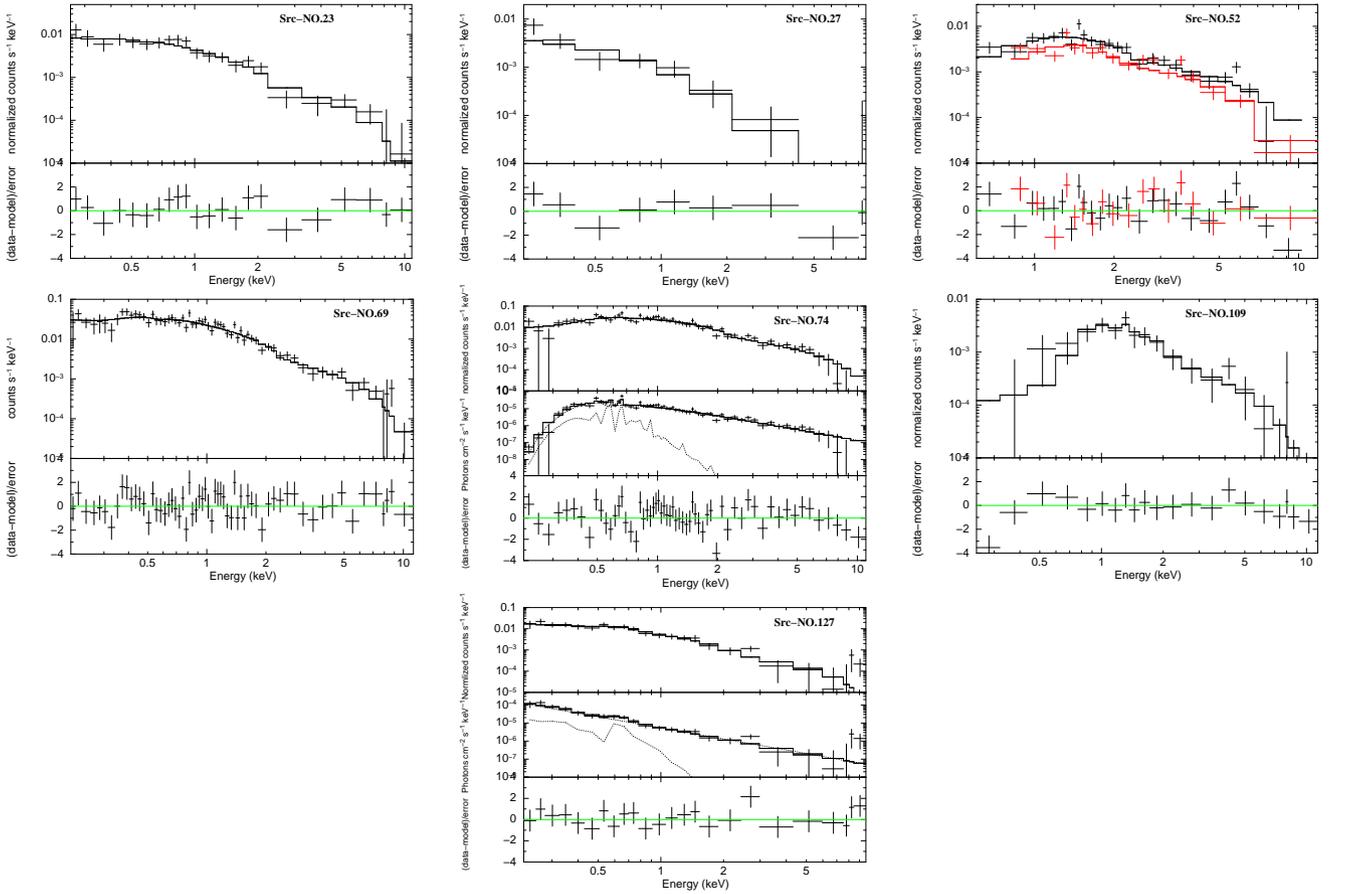

\includegraphics[angle=270, width=0.33\textwidth, trim=1.5cm 0.cm 0.cm 2.0cm]{23.eps}
\includegraphics[angle=270, width=0.33\textwidth, trim=1.5cm 0.cm 0.cm 2.0cm]{source-combined_27.eps}
\includegraphics[angle=270, width=0.33\textwidth, trim=1.5cm 0.cm 0.cm 2.0cm]{52.eps}\\
\includegraphics[angle=270, width=0.33\textwidth, trim=1.5cm 0.cm 0.cm 2.0cm]{source-combined_69-new.eps}
\includegraphics[angle=270, width=0.33\textwidth, trim=1.5cm 0.cm 0.cm 2.0cm]{source-combined_74_unfold.eps}
\includegraphics[angle=270, width=0.33\textwidth, trim=1.5cm 0.cm 0.cm 2.0cm]{109.eps}\\
\includegraphics[angle=270, width=0.33\textwidth, trim=1.5cm 0.cm 0.cm 2.0cm]{source-combined_127_unfold.eps}
\caption{Combined \xmm\, spectra of X-ray sources  \label{spec.fig}} 
\end{figure*}

We analysed the X-ray spectra of the  bright sources in the field of Sculptor dSph, which have not been classified as background objects (see Sect.~\ref{diss}) and had total net source counts of >500. By relying on the method of our previous works \citep{2019A&A...627A.128S}, we improved the statistics of the spectra by merging the spectra of both observations, in case that the source was detected in both observations, using the \texttt{SAS} task \texttt{epicspeccombine}. Before merging the spectra of different observations, the sources were checked for variability. Spectra of sources, which showed a significantly different flux in the two observations (see table~\ref{catalogue-x-ray}) were not merged. Figure~\ref{spec.fig} shows the spectra and Table~\ref{spectral-Table} the parameter values of the best-fit models for the spectra of the sources. For the spectral analysis we  used the X-ray absorption model \citep[\texttt{tbabs},][]{2000ApJ...542..914W}, ionised plasma model for the soft X-ray emission \citep[\texttt{apec},][]{2001ApJ...556L..91S} and power law model for the hard X-ray emission. The model for each source is selected based on the best fit and are shown in Fig.~\ref{spec.fig} and Table.~\ref{spectral-Table}. The details of the spectral study of each source are discussed in Sect.~\ref{diss}. The HR study is considered as an alternative way to determine the properties of the spectrum of the sources, which were too faint for a spectral analysis (see Sect.~\ref{hr-sec}). 
\begin{table*}
\centering
\caption{Best-fit parameters of the X-ray spectra. Errors are at the 90$\%$ confidence level. \label{spectral-Table}}
\small
\centering
\addtolength{\tabcolsep}{-0.1cm}   
\begin{tabular}{llccccccc}
\hline\hline
Src-No & Model & $N_{\rm H}$&Photon index& $kT$ & Abundance&$\chi^2$ (d.o.f) &  Absorbed $F_{\rm X}$&$L_{\rm X}^{(1)}$\\
&&$10^{22}$ cm$^{-2}$&&keV&&&$10^{-14}$erg\,s$^{-1}$\,cm$^{-2}$&erg\,s $^{-1}$\\
\hline
\hspace{2mm}
23&\texttt{tbabs$\times$(po)}& <0.07&$1.7^{+0.45}_{-0.16}$&&& 1.2 (20) &$3.05^{+0.34}_{-0.34}$& $3.0\times10^{34}$\\
\hspace{2mm}
27&\texttt{tbabs$\times$(po)}& <0.07&$2.42^{+0.82}_{-0.53}$&&&  1.5(6) &$0.3^{+0.08}_{-0.07}$& $2.6\times10^{33}$\\
\hspace{2mm}
52&\texttt{tbabs$\times$(po)}& 0.018 frozen &$1.73^{+0.23}_{-0.21}$&&&1.2(39) &6.26$^{+0.76}_{-0.73}$&6.2  $\times10^{34}$\\
&  &$0.40^{+0.14}_{-0.12}$ &&&&&&\\
\hspace{2mm}
69&\texttt{tbabs$\times$(po)}& $0.03^{+0.02}_{-0.02}$ &$1.83^{+0.15}_{-0.13}$&&&1.02(61) &$10.81^{+0.15}_{-0.15}$&9.5 $\times10^{34}$\\
\hspace{2mm}
74&\texttt{tbabs$\times$(apec+po)}& $0.16^{+0.08}_{-0.05}$& $2.10^{+0.23}_{-0.20}$ &$0.18^{0.06}_{-0.06}$ & & 1.26 (55) &$6.91^{+0.05}_{-0.04}$ &$6.1\times10^{34}$\\
\hspace{2mm}
 109& \texttt{ tbabs$\times$(po)}& $0.44^{0.32}_{-0.21}$& $2.37^{+1.04}_{-0.68}$ & & & 1.23 (22) & $1.2^{+0.70}_{-0.75}$&8.2$\times10^{33}$\\
\hspace{2mm}
127&\texttt{tbabs$\times$(po+apec)}&<0.06& $2.03^{+0.33}_{-0.29}$& <0.7 & &1.18(20)&$3.14^{+0.20}_{-0.22}$& 3.1$\times10^{34}$\\
\hline
\multicolumn{9}{l}{The distance of the Sculptor dSph of 84 kpc was used to calculate the luminosity of the sources.}
\end{tabular}
 \end{table*}

\section{Multi-wavelength studies of counterparts}
\label{multi}
In the following, we discuss the multi-wavelength photometry used to uncover the  nature of our sources. 
\subsection{Optical counterparts of the sources}
\label{multi-opt}
The most updated survey in the field of Sculptor dSph is the first data release of the dark energy survey \citep[DES,][]{2018ApJS..239...18A}. The catalogue includes photometric data in the energy bands from the optical to the near infrared ($g$=400--500~nm, $r$=600--750~nm, $i$=750--1000~nm , $z$=750--1000~nm, $Y$=1000--1500~nm) and allows a spectral study of the optical counterpart. For the $g$, $r$, $i$, $z$, and $Y$ bands, the Galactic extinction of 0.04, 0.03, 0.02, 0.02, and 0.01 mag is used in the direction of the Sculptor, respectively \citep{2011ApJ...737..103S}. Table~\ref{opt-count-table} lists the DES magnitudes of the optical counterparts of the X-ray sources.

\begin{figure*}
\centering
\includegraphics[trim={3.2cm 1.0cm 0.0cm 0.cm},width=0.45\textwidth]{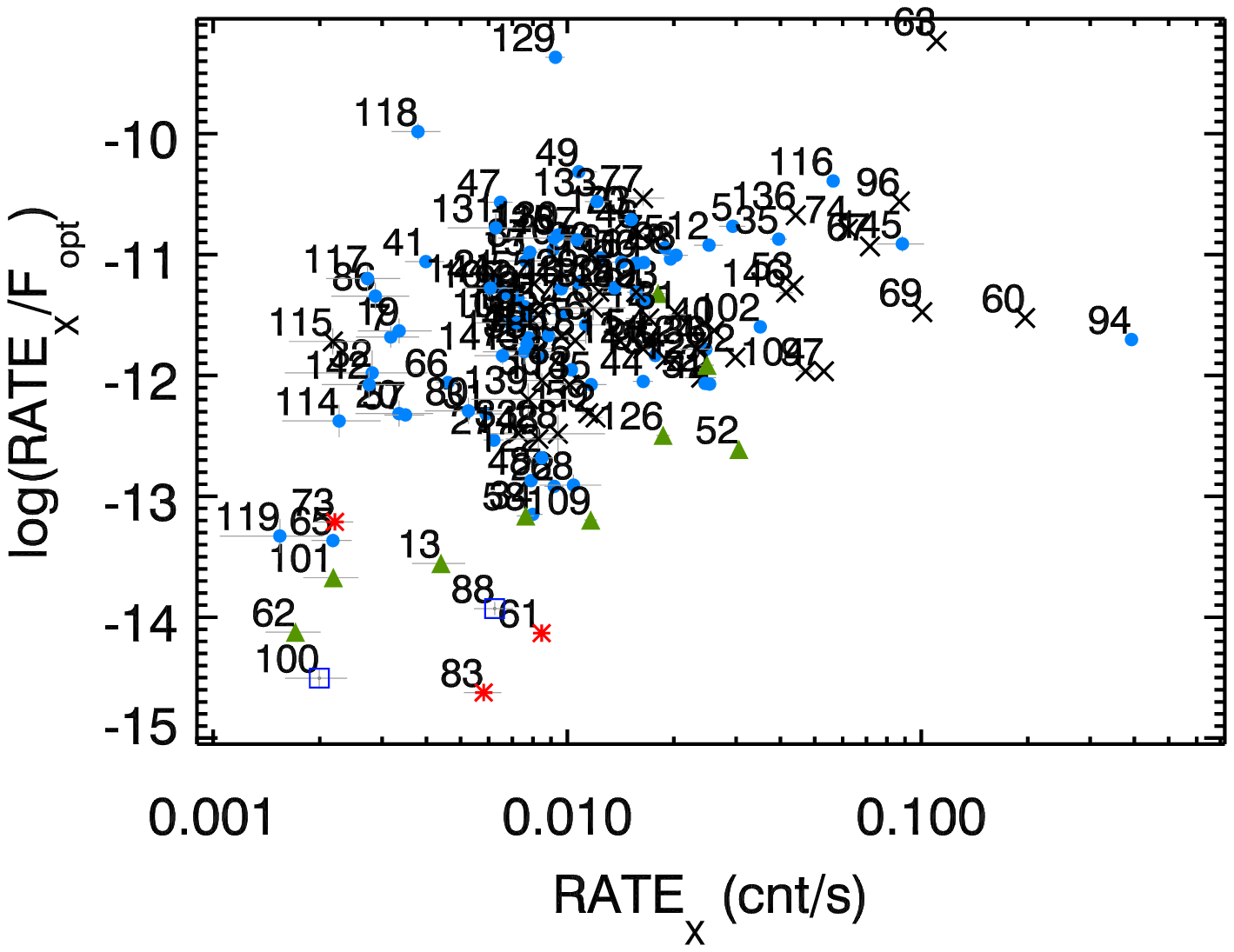}
\includegraphics[trim={3.2cm 1.0cm 0.0cm 0.cm},width=0.45\textwidth]{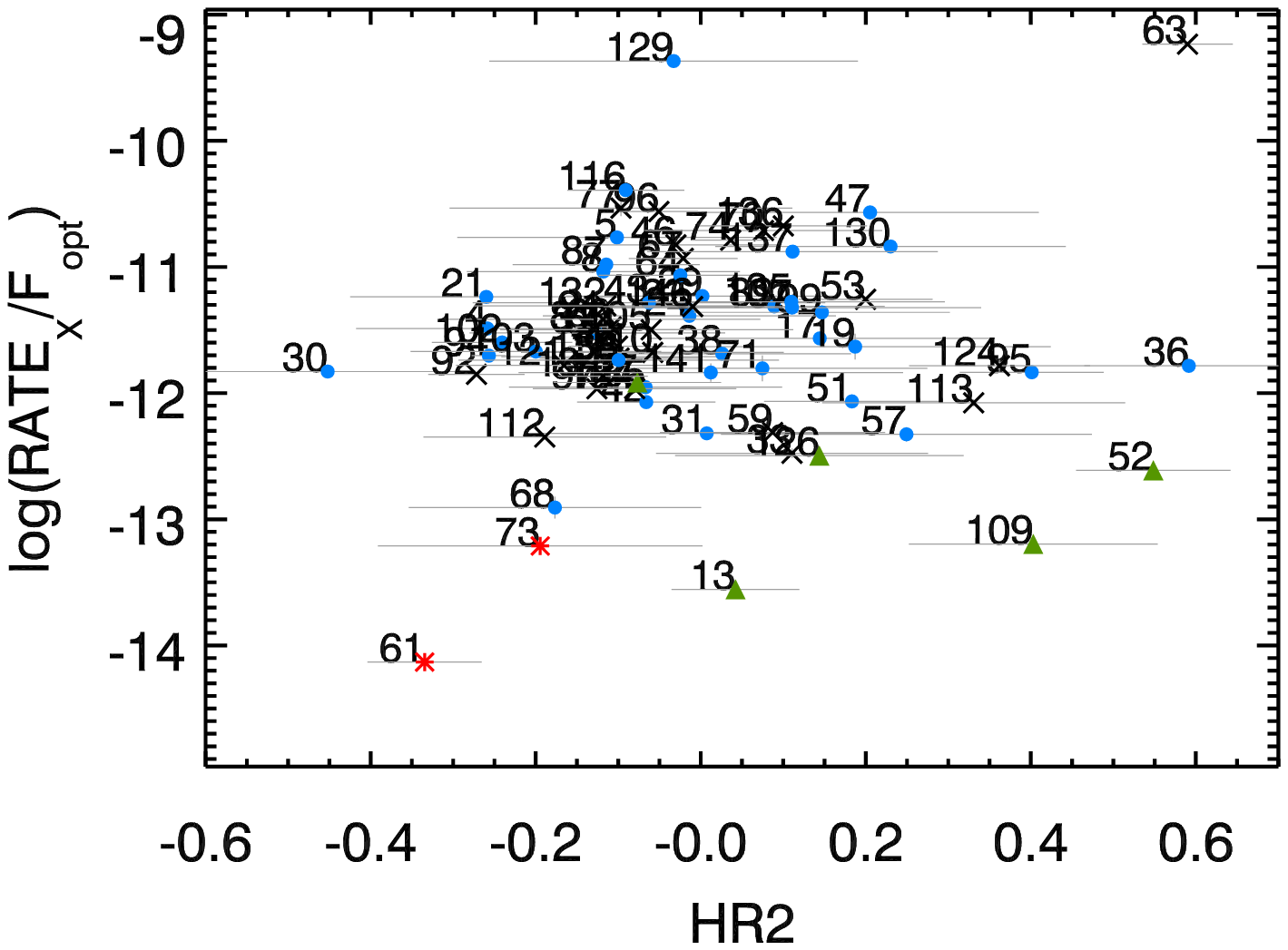}
\caption{Logarithmic X-ray rate to optical flux ratio log$(\frac{RATE_\text{X}}{F_\text{opt}})$ over the maximum X-ray count rate (left panel) and over the HR2 (right panel). The symbols are the same as in Fig.~\ref{var}.\label{log-x-opt}}
\end{figure*}

For comparing the fluxes of a source in the optical and in the X-ray bands, we calculate the logarithm of the ratio of the fluxes.
Since the flux of the X-ray sources depends on the selected model, we decided to modify the X-ray to optical flux ratio equation  of  \citet{1988ApJ...326..680M} and use the X-ray rate. This has the advantage that the X-ray rates represent the brightness of the X-ray sources directly and is not subject to the uncertainties of the ECFs.
The equation thus becomes
\begin{equation}
{\rm log}\bigg(\frac{Rate_\text{X}}{F_\text{opt}}\bigg)=\rm{log}_{10}(F_\text{X})+\frac{g+r}{2\times2.5}+5.37
\end{equation}

where $RATE_\text{X}$ is the X-ray rate and $g$ and $r$ are the magnitudes of the optical counterpart of the X-ray source. We plotted  log$(\frac{RATE_\text{X}}{F_\text{opt}})$ versus the X-ray rate to be able to check the brightness of X-ray sources in comparison to their optical brightness, as well as  log$(\frac{RATE_\text{X}}{F_\text{opt}})$  versus HR2 to see the correlation between the spectrum and the X-ray-to-optical brightness ratio of the sources (see Fig.~\ref{log-x-opt}). 

\begin{figure}
\centering
\includegraphics[trim={3.2cm 1.0cm 0.0cm 0.cm},width=0.47\textwidth]{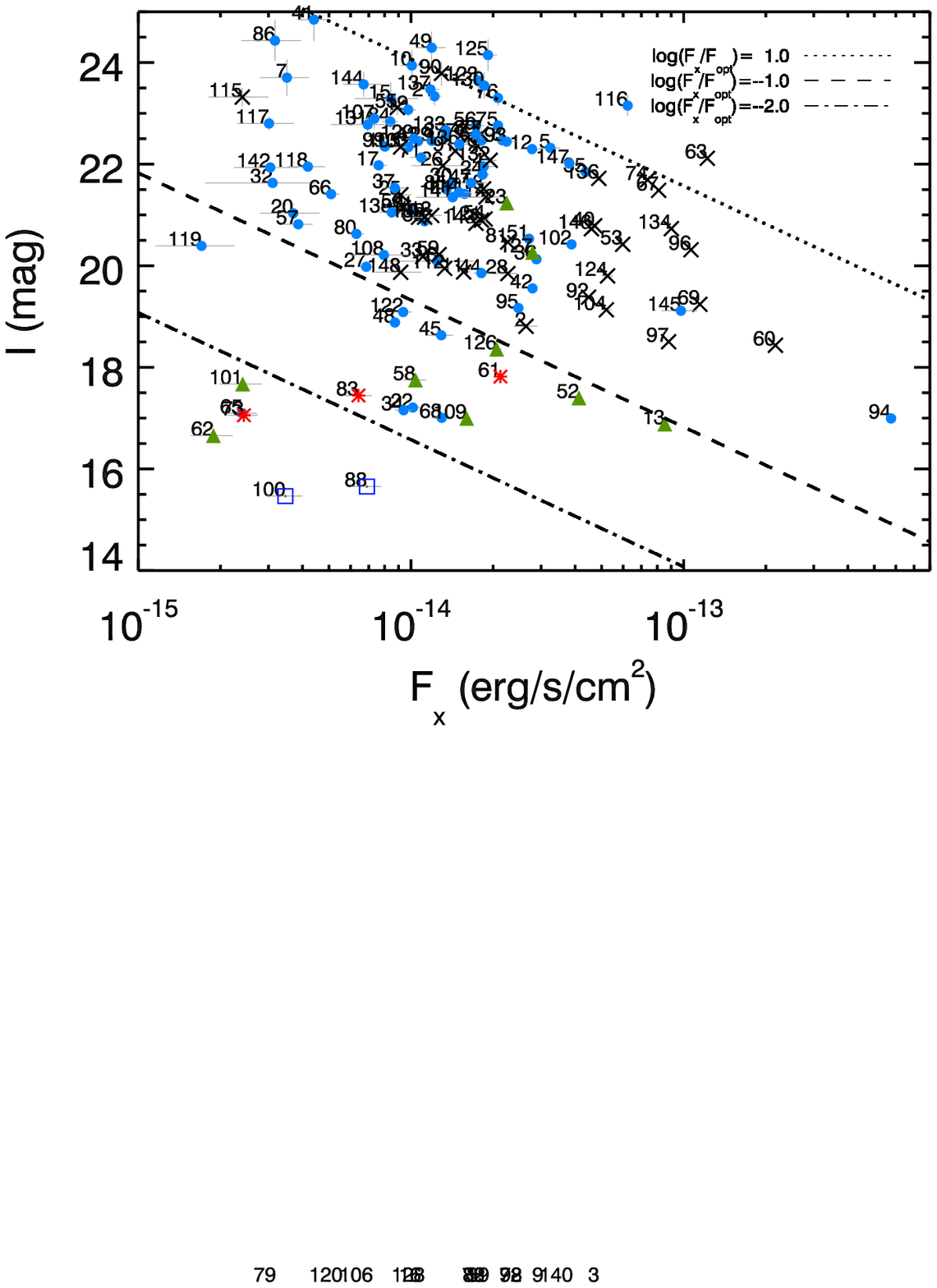}
\caption{ X-ray flux versus I-band magnitude of DES counterpart of the sources.  The lines show the constant logarithmic X-ray rate-to-optical flux ratio log$(\frac{F_\text{X}}{F_\text{opt}})$. The symbols are the same as in Fig.~\ref{var}.\label{imag-flux}}
\end{figure}

 Using the optical counterparts of the X-ray source, we plotted the $I$ magnitude versus the X-ray flux (see Fig.~\ref{imag-flux}) similar to the study of \citet{2002ApJ...568L..85A}, in which the behaviour of background X-ray sources has been studied using the optical counterparts.  \citet{2002ApJ...568L..85A} have shown that there are regions in the plot, which are dominated by  AGNs (--1.0<log$(\frac{Flux_\text{X}}{F_\text{opt}})$)<1.0), starburst galaxies (--2.0<log$(\frac{Flux_\text{X}}{F_\text{opt}})$)<--1.0), and normal galaxies (log$(\frac{Flux_\text{X}}{F_\text{opt}})$)<--2.0). The results are discussed in Sect.~\ref{bk-diss}. 
\begin{figure}
\centering
\includegraphics[trim={3.2cm 0.75cm 0.0cm 0.cm},width=0.45\textwidth]{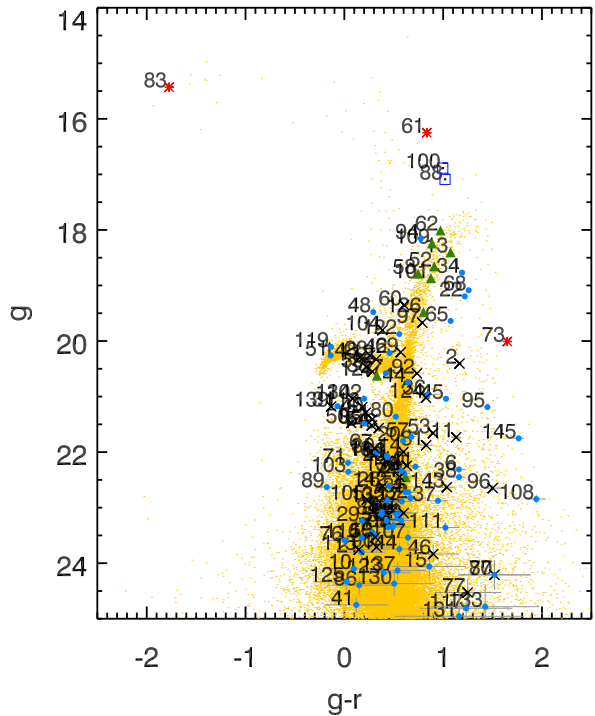}
\caption{The optical counterparts of the X-ray sources in the field of Sculptor dSph. The yellow dots are all optical sources detected in the first data release of DES \citep{2018ApJS..239...18A} in the field of Sculptor dSph. The symbols are the same as in Fig.~\ref{var}.  \label{opt-counterpart} }
\end{figure}

\subsection{Infrared counterparts of the sources}
\label{infra-count-sect}
\begin{figure}
\centering
\vspace{-0.75cm}
\includegraphics[trim={3.2cm 0.0cm 0.0cm 0.cm},width=0.45\textwidth]{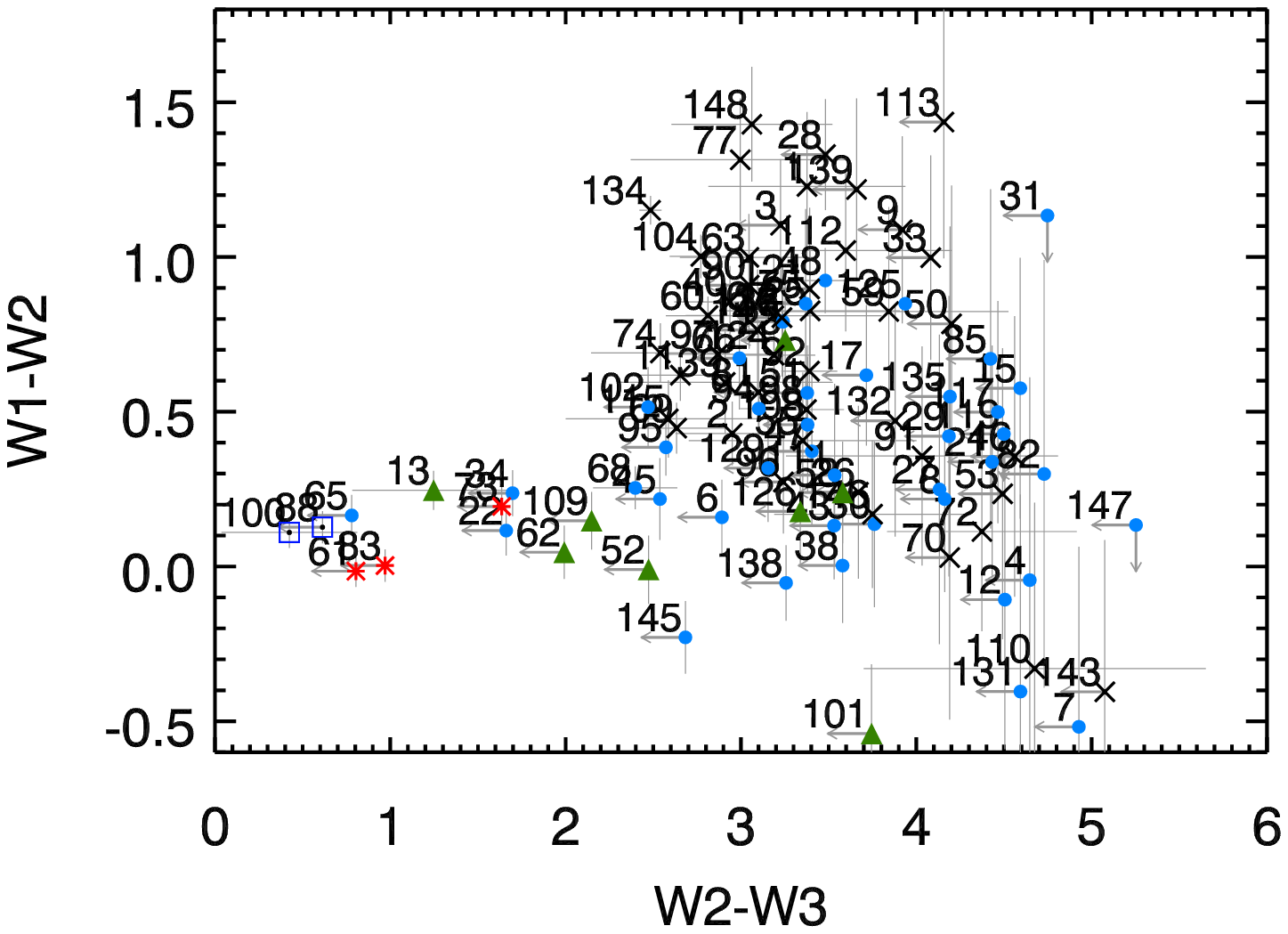}
\includegraphics[trim={3.2cm 1.0cm 0.0cm 1.1cm},width=0.45\textwidth]{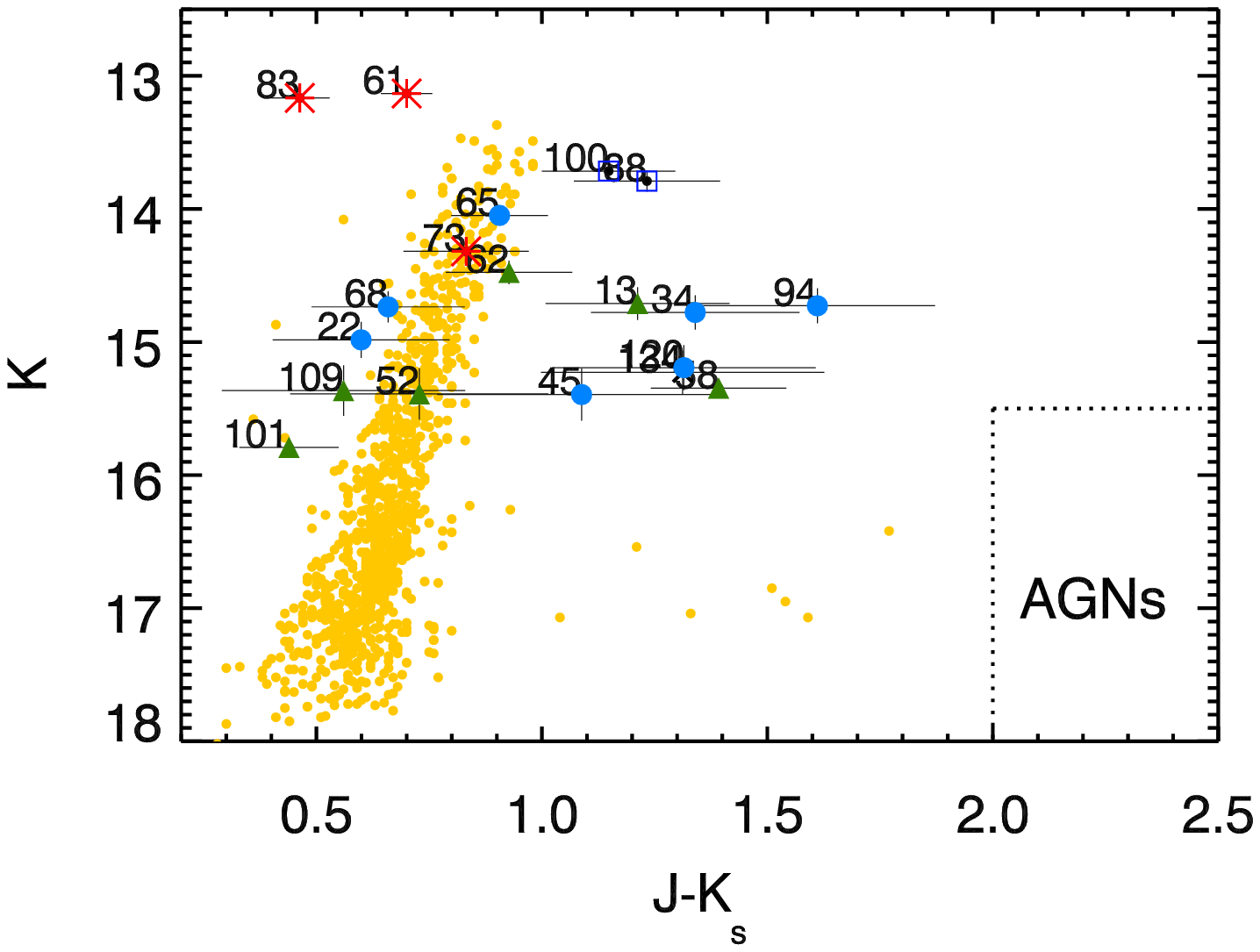}
\caption{{\it Upper panel:} Colour-colour diagram of mid-infrared WISE  ($W1(3.4\, \mathrm{\mu m})-W2( 4.6\, \mathrm{\mu m})$ versus ($W2\,(4.6\, \mathrm{\mu m})- W3 \,(12\, \mathrm{\mu m})$). Arrows show  upper limits (see Table~\ref{inf-count}). {\it Lower panel:} Colour magnitude diagram of 2MASS counterparts of the X-ray sources, which have not been located in the AGN region defined by \citet{2010PASA...27..302M}. The yellow points shows the main population of Sculptor dSph in 2MASS catalogue. The symbols are the same as in Fig.~\ref{var}. \label{infra-plot}}
\end{figure}

We searched for mid-infrared counterparts in the WISE All-Sky survey in four energy bands \citep[3.4, 4.6, 12, and 22 $\mathrm \mu$m, named $W1$, $W2$, $W3$, and $W4$, respectively;][]{2014yCat.2328....0C}. 
The Galactic extinction for the infrared bands in the direction of Sculptor dSph was negligible \citep{2011ApJ...737..103S}.  
 For the study of near-infrared counterparts, we use the 2MASS All-Sky Catalogue of Point Sources in the $J$, $H$, $K$ bands \citep{2003yCat.2246....0C}. 
Table~\ref{inf-count} lists the magnitudes of WISE and 2MASS counterparts of the X-ray sources.  Figure~\ref{infra-plot} shows the colour-colour diagram of the WISE counterparts of the detected X-ray sources. According to  \citet{2010AJ....140.1868W}, usually, background sources (AGN or galaxies) are expected to be red ($W2-W3$>1.5) in WISE colour, while stellar sources show lower $W2-W3$<1.5.  Based on 2MASS near-infrared photometry, background sources like AGNs and quasars  are expected to have $J-K>1$ \citep[see e.g,][]{2010PASA...27..302M}. 
Following \citet{2010PASA...27..302M}, we tried to distinguish the stellar sources from the AGNs based on 2MASS photometry. For AGNs, the criteria are: colour $J-K_{s}$>2.0, magnitude $K_{s}$>15.5, and the source is located at a Galactic latitude of $|b|$>30$^{\circ}$.
For red giant\,(RG) stars in the field of sculptor dSph, we used the criteria reported by  \citet{2011MNRAS.414.3492M}. Figure\,\ref{infra-plot} (lower panel) shows the colour-magnitude diagram of near-infrared counterparts of the 2MASS catalogue.

 In addition, for the classification of symbiotic stars, we used the results of the machine learning method of \citet{2019MNRAS.483.5077A}. The main population of symbiotic stars have  $J-H>0.78$ and only a small fraction of S-type symbiotics behaves differently. The second criteria is  $K-W3<1.18$, which helps to distinguish the symbiotic stars from the other types of sources. However, for  dusty symbiotic stars the second criterion might not work. In this case, there are two other criteria for finding symbiotic stars: $H-W2>3.80$ and $W1-W4<4.72$. We tried to apply these criteria, however, in many cases the counterparts were faint and had large errors and/or only upper-limits in some infrared energy bands, which made a clear classification difficult. The details are explained in Sect.~\ref{diss}. 

\subsection{Catalogues of AGNs and galaxies}
\label{AGN-cata}
For the classifications of AGNs and background galaxies besides the colours of infrared counterparts, the following catalogues were considered as well: the Million Quasars catalogues \citep{2019yCat.7283....0F}, a compilation of known quasars for the Gaia mission \citep{2019RAA....19...29L}, the quasars and galaxy classified using Gaia data \citep{2019MNRAS.490.5615B}, star, galaxy, and quasars classification of \citet{2020A&A...633A.154L}, and the catalogue of quasars identified with astrometric and mid-infrared methods from  the  catalogue of the Absolute Proper motions Outside the galactic Plane (APOP) and WISE \citep{2018A&A...618A.144G}.
\subsection{ Gaia and GALEX catalogues}
To classify the foreground sources, we used the photometry and parallax measurements of the 3rd data release of Gaia \citep{2020yCat.1350....0G} and the GALEX catalogue of UV  sources, which includes Galactic white dwarfs and hot stars \citep{2017ApJS..230...24B}.

\subsection{Catalogues of stars in Sculptor dSph}
\label{sculp-cata}
There are several infrared and optical studies, which have classified the members of Sculptor dSph based on radial velocity, metallicity, and photometry. For all X-ray sources, we checked if the counterpart has been confirmed as a Sculptor member. The following studies have been used for our work:

\begin{itemize}
\item Catalogue of the absolute proper motion and a membership probability of the Sculptor dSph extracted  from  26 optical available surveys \citep{1995AJ....110.2747S},
\item Catalogue of stellar velocities of some dSphs (including Sculptor) using the data of the Michigan/MIKE Fiber System Survey\,(MMFS) \citep[][]{2009AJ....137.3100W},
\item Catalogue of asymptotic giant branch stars in the Sculptor dSph using the infrared data of Japanese South African Infrared Survey Facility (IRSF) \citep{2011MNRAS.414.3492M},
\item Catalogue of universal Stellar mass-Stellar metallicity relation for dSphs using the data of  Keck Telescopes \citep{2013ApJ...779..102K}, and
\item Catalogue of the study of Zinc abundances in the Sculptor dSph based on a spectroscopy study with the FLAMES/GIRAFFE cameras of the Very large telescope\,(VLT) \citep{2017A&A...606A..71S}.
\end{itemize}

\section{Discussion}
\label{diss}
\subsection{Background sources} 
\label{bk-diss}
For the classification of the background sources we relied on the classification of other available catalogues  as discussed in Sect.~\ref{AGN-cata}. Moreover, if a source has a WISE counterpart with significant magnitudes in $W1$, $W2$, and $W3$ bands, background objects were separated from the stellar sources as explained in Sect.~\ref{infra-count-sect}.  As Fig.~\ref{log-x-opt} shows
AGNs are bright both in X-rays and in the optical and thus have higher log$(\frac{RATE_\text{X}}{F_\text{opt}})$. The study of optical $I$-band magnitude versus the X-ray flux can provide a hint about the nature of background sources as we discussed in Sect.~\ref{multi-opt}. As Fig.~\ref{imag-flux} shows, all classified AGNs  are in the region, where --1.0<log$(\frac{Flux_\text{X}}{F_\text{opt}})$<1.0 (Sect.~\ref{multi-opt}). Source No.\,48 and No.34 are in the region of starburst galaxies and source No.\,88 and No.\,100 are normal galaxies. We also discuss  these two optically bright sources in Sect.~\ref{can-diss}. 

In total, we classified 46 sources as background objects in the field of Sculptor dSph. One should note that the expected number of background sources according to the log\,N--log\,S  AGN distribution \citep{2009A&A...497..635C} in the energy band of 0.5--2.0 keV for the field of view of the EPIC-pn camera (0.154\,deg$^{-2}$) is $\sim$90 AGNs for a flux >$1.7\times10^{15}$~erg\,s$^{-1}$\,cm$^{-2}$. Not all of the background AGNs could be identified due to the lack of information necessary for source classification.

\subsection{Foreground stars} 
Our classification suggests only three sources to be foreground stars in the Milky~Way. The low number of  foreground stars in the field of Sculptor dSph is not surprising due to its location near the Galactic South pole, with a short line of sight through the the Galactic disk.  Fig.~\ref{log-x-opt} shows that foreground stars  have low log$(\frac{RATE_\text{X}}{F_\text{opt}})$ compared to background sources. They are X-ray faint (Fig.~\ref{log-x-opt}, left) and have a soft spectrum (Fig.~\ref{log-x-opt}, right).
We will quickly present each source in the following.

{\bf Source No.\,61:} It is a bright Galactic foreground star already discussed in \citet[X-12,][]{2019MNRAS.485.2259A}.

{\bf Source No.\,73:} It is a stellar object according to the  WISE infrared colour and has a parallax of 290$\pm$12 pc in the Gaia catalogue \citep{2020yCat.1350....0G}. Considering the optical and the 2MASS colours    \citep[$i-z$, $z-J$,$J-H$, and $H-K_{s}$,][]{2011AJ....141...97W}, the source is a candidate for a M4-type dwarf. The distance of the Gaia counterpart (290$\pm$13 pc) suggests a luminosity of $\sim2.4\times10^{28}$~erg\,s$^{-1}$ for the source. 

{\bf Source No.\,83:} The infrared counterpart is a stellar object candidate (see Fig.\ref{infra-plot}). The infrared (2MASS) and optical colour magnitude diagrams show that it is brighter than the population in Sculptor dSph and thus should be a foreground source (see Fig.~\ref{infra-plot} and Fig.~\ref{opt-counterpart}).  Gaia survey measured a distance of $808\pm28$~pc for the star. The source was too faint and outside the field of view of EPIC pn in the second observation, therefore, no information about the spectrum and HRs is available. 

\subsection{Candidates of X-ray sources in the Sculptor dSph}
\label{can-diss}
The study of X-ray sources by \citet{2019MNRAS.485.2259A} focused on thirteen X-ray sources: five sources, which have been classified as LMXB candidates in the previous study using \chandra\, observations \citep{2005MNRAS.364L..61M} and  eight other bight sources observed with \chandra\,(OBS ID: 9555, 2009, with ACIS-S). In the following, we discuss  the sources, which  have been observed with \xmm\ and their optical/infrared counterparts suggest them to be a member of Sculptor dSph. Moreover, we  discuss  the sources, which were analysed by \citet{2019MNRAS.485.2259A} and their nature was not fully clear. 

{\bf Source No.\,13:} The counterpart of the source is a stellar object according to the WISE colours (see Fig.\ref{infra-plot}) and in the optical, it is located at the tail of the red-giant branch\,(RGB) 
of Sculptor (see Fig.~\ref{opt-counterpart}). In the absolute proper motion measurement of \citet{2015AJ....150..137Q}  the source was also classified as an extra-galactic star. It makes the source most probably a member of Sculptor dSph.
The source shows neither pulsation nor periodicity. It was too faint for the spectral analysis. However, the HRs of the source show that its count rate decreases significantly >4.5 keV (see Table \ref{catalogue-x-ray} and Fig.\ref{hrs-plot}). The luminosity of the source assuming the distance of Sculptor dSph is of the order of 10$^{33}$~erg\,s$^{-1}$.  Using the infrared colour criteria of \citet{2019MNRAS.483.5077A} (see Sect.~\ref{infra-count-sect}) the colours of the infrared counterpart of the source ($J-H=0.77\pm0.19$, $K-W3=2.10\pm0.55$) suggest that it is not a symbiotic star. Colours of  $H-W2=1.29\pm0.14$ and $W1-W4<4.86$ give a low probability to the source to be a dusty symbiotic star. Therefore, the X-ray source has a low chance to have this red giant as its companion and the source remains unclassified.

{\bf Source No.\,22:} The source shows WISE and 2MASS magnitudes similar to those of RGB stars in the Sculptor dSph (see Fig.~\ref{infra-plot}). However, the optical counterpart  of the source is not  on the RGB of the Sculptor dSph (see Fig.~\ref{opt-counterpart}). Available catalogues confirm  the counterpart of the source neither as an extra-galactic star \citep{2015AJ....150..137Q} nor as a member of Sculptor dSph \citep{2011MNRAS.414.3492M}.  The infrared counterpart does not have significant values to make it clear if the companion is a red giant in the symbiotic star system (see Sect.~\ref{infra-count-sect}). The low statistics did not allow to perform a spectral analysis. No pulsation is detected for the source. Therefore, the source remains unclassified. 

{\bf Source No.\,23:} The study of \citet{2016MNRAS.462.4349M} has considered the counterpart of this source as a candidate for a variable star in Sculptor dSph. The optical counterpart of the source is located on the main sequence of the Sculptor dSph (see Fig.~\ref{opt-counterpart}). The X-ray spectral analysis shows that it mainly emits below 10.0~keV and has a luminosity of $\sim10^{34}$~erg\,s$^{-1}$. The luminosity of QLMXBs can be $\sim10^{32}-10^{34}$~erg\,s$^{-1}$ and their main X-ray emission is <5.0~keV \citep[e.g,][]{2003ApJ...598..501H, 2000ApJ...539..191Y, 2019MNRAS.487.2005C}. The luminosity, X-ray spectrum,  and  the main sequence stellar counterpart make this source a candidate for a QLMXB. 

{\bf Source No.\,27:} The optical counterpart is classified as a member of the Sculptor dSph according to its stellar velocity \citep{2009AJ....137.3100W}. As seen in the optical-colour magnitude diagram, the source is located on the horizontal branch of Sculptor dSph (see fig.~\ref{opt-counterpart}).  Its X-ray spectrum has too poor statistics to be fit with a model, however, the combined spectrum of both observations shows that the source has  no emission above >5.0 keV (Fig.~\ref{spec.fig}). No pulsation has been found for the source. The flux of 1.8$\times$10$^{-15}$~erg\,cm$^{-2}$\,s$^{-1}$ (luminosity of the order of $\sim$10$^{33}$~erg\,s$^{-1}$ at the distance of Sculptor dSph) suggests that it can be  a candidate for a faint QLMXB, however its nature can not be determined ultimately.

{\bf Source No.\,52:} The study of \citet[source X-11,][]{2019MNRAS.485.2259A} suggests no counterpart for the source. However, the deeper \xmm\, observations of our work suggest a position for the source with a bright optical/infrared source at $\sim$1.8$\arcsec$ offset (see Fig.~\ref{opt-image}), while the \chandra\, X-ray position of the source is $\sim$2.9$\arcsec$ away from the same optical/infrared source.  According to the WISE colours the counterpart is a stellar object (see Fig.\ref{infra-plot}). The optical counterpart is located on the RGB of the Sculptor dSph (see Fig.~\ref{opt-counterpart}). In several studies the infrared/optical counterpart of this source is classified as a member of Sculptor dSph based on  photometry \citep{2009AJ....137.3100W, 2011MNRAS.414.3492M}, abundance \citep{2013ApJ...779..102K}, and  radial velocity  \citep{2009AJ....137.3100W, 2017A&A...606A..71S}. The colour of the companion of the source in infrared is $J-H=0.65\pm0.21$. This colour error is too large to verify the first criteria ( $J-H>0.78$, see Sect.~\ref{infra-count-sect}) for the classification of typical symbiotic stars. We can not check the other infrared criteria either, due to the upper-limits in the energy bands. If we assume the RG as the companion of the source, the X-ray luminosity ($\sim6.\times$10$^{34}$~erg\,s$^{-1}$) makes it  a candidate for a $\gamma$-type symbiotic star (consisting of a neutron star and an RG companion) or  a non-magnetic white dwarf symbiotic star with dominating emission from the boundary layer. The spectral analysis shows a hard spectrum with a high intrinsic absorption for the source, which might be related to the boundary layer of the compact object. The source looks similar to  XMMUJ171919.8+575943 (Src-No.\,1) in Draco dSph studied by \citet{2019A&A...627A.128S}.  We classify this source as a symbiotic star candidate in the Sculptor dSph.

{\bf Source No.\,58:} The infrared counterpart of the source is classified as a member of Sculptor \citep{2011MNRAS.414.3492M}.  Abundance and radial velocity studies have also confirmed that the source counterpart is a member of Sculptor dSph \citep{2013ApJ...779..102K, 2009AJ....137.3100W, 2017A&A...606A..71S}. The optical colour magnitude diagram shows that the source is located on RBG of Sculptor dSph (see Fig.~\ref{opt-counterpart}). The main infrared colour of the counterpart, which can tell if the source  is a symbiotic star has very large error ($J-H=0.97\pm0.31$), and thus does not allow to clarify if the source is a symbiotic star (see Sect.~\ref{infra-count-sect}).  The X-ray source was too faint for its spectrum to be studied.  The hardness ratio is only significant for HR1 (see Table~\ref{catalogue-x-ray}), which means that the main emission of the source is  soft. The luminosity  at the distance of Sculptor dSph is $\sim$9$\times 10^{32}$~erg\,s$^{-1}$ and suggests that it is most probably a candidate for a symbiotic star.

{\bf Source No.\,62:} The counterpart of the source is already classified as an RG \citep{2001MNRAS.327..918T} and a member of Sculptor dSph \citep[][]{1995AJ....110.2747S, 2009AJ....137.3100W}.  The source was too faint for its spectrum and HRs to be analysed. The luminosity of the source is $\sim10^{33}$~erg\,s$^{-1}$.  2MASS infrared counterpart of the source has a colour of $J-H=0.66\pm0.11$, which is slightly lower than the criterion ($J-H\geq$ 0.78, see Sect.~\ref{infra-count-sect}), therefore it is not clear if an RG is the companion of this X-ray source and thus the source stays unclassified. 

{\bf Source No.\,68:} The counterpart of the source is not confirmed as an RG in available studies/catalogues. The WISE infrared counterpart is  closer to the colour  of background galaxies ($W2-W3>1.5$, see Fig.~\ref{infra-plot}). Moreover, in the optical colour magnitude diagram  it is not located on the isochrone of RGBs of the Sculptor dSph (see Fig.\ref{infra-plot}).  The source was too faint for a spectral analysis and its hardness ratios show that the main emission is observed below 2.0~keV (see Table~\ref{catalogue-x-ray}). It is most likely a background galaxy candidate, however this can not be fully confirmed.

{\bf Source No.\,69:} The details of the optical spectrum of the source has been discussed in \citet[][source SD X-1]{2019MNRAS.485.2259A}. As \citet{2019MNRAS.485.2259A} point out, the earliest studies  suggested that the source is a blazer. However, the Gemini Multi-Object Spectrographs\,(GMOS) spectrum absorption lines are not consistent with this scenario. It has been discussed that  the source is either a blazer and the  absorption lines come from its emission passing through the interstellar medium in the Sculptor dSph, or there might be  a  star in the Sculptor dSph with absorption lines caused by this star. The X-ray spectral analysis of the source shows a very low column density almost comparable the Galactic absorption in the direction of Sculptor dSphs (see Fig.~\ref{spec.fig} and Table~\ref{spectral-Table}). Therefore, we believe that a high absorption in the Sculptor dSph is less likely. However,  the nature of this source stays unclear. 

{\bf Source No.\,74:} The source is considered as a possible QLMXB in  the work of \citet[][source SD X-5]{2019MNRAS.485.2259A}. We have performed  X-ray spectral and timing analyses to improve our knowledge about the source. The absorption is slightly higher than the Galactic absorption in the direction of Sculptor dSph. The spectrum has a soft thermal component and a power-law model component, which is fit well to the higher energy tail of the emission. It seems to be softer than that of typical AGNs \citep{2011A&A...530A..42C} (see Table~\ref{spectral-Table}), however, since the WISE colour of the  counterpart of the source is consistent with that of  AGNs/galaxies (see fig.~\ref{infra-plot}) it can not be confirmed as a QLMXB. As already mentioned by \citet{2019MNRAS.485.2259A}, it is most probably a background AGN/galaxy. 

{\bf Source No.\,88 and No.\,100:} Even though the infrared colour of these two sources are similar to that of the stellar objects, the  optical counterpart of them look very bright and are located above the RGB of Sculptor (see Fig.~\ref{inf-count} and Fig.~\ref{opt-counterpart}). Studies confirm that these sources are neither a  Galactic stellar objects \citep{2020yCat.1350....0G} nor a members of the Sculptor (see Sect~\ref{sculp-cata}).   \citet{2015AJ....150..137Q} classified these sources as a non-stellar extra-galactic object according to the proper motion.  The comparison between the $i$-band optical magnitude and the X-ray flux shows that the source in the region of normal galaxies (see Fig.~\ref{imag-flux} and Sect.~\ref{multi-opt})   \citet{2020A&A...633A.154L} classified the sources as galaxy candidates. Therefore, we considered these two sources as background galaxies.

{\bf Source No.\,101:} The optical and infrared counterparts of the source are classified as an RG in the Sculptor dSph \citep{2017A&A...606A..71S, 1995AJ....110.2747S, 2011MNRAS.414.3492M}. The source was only in the field of view of EPIC-pn of OBS1 and too faint to be detected with EPIC-MOS. The spectral analysis and HR study of the source was not possible due to the very low statistics. The X-ray luminosity assuming to be at the distance of Sculptor dSph, is  $\sim2 \times10^{33}$~erg\,s$^{-1}$.  2MASS infrared colour of the source as the main criterion of being a symbiotic star is $J-H=0.36\pm0.28$, which is significantly different from that of the known symbiotic stars (see Sect.~\ref{infra-count-sect}). Therefore the RG counterpart can not be related to this X-ray sources and the source remains unclassified. 

{\bf Source No.\,109:} The source counterpart is already confirmed as a member of Sculptor dSph \citep{2009AJ....137.3100W,2011MNRAS.414.3492M}. The optical colour magnitude diagram shows that it is in the RGB of  Sculptor dSph (see Fig.~\ref{opt-counterpart}). The infrared counterpart of the source has very large errors, which do not allow to check if the source meets the symbiotic star colour criteria (see Table~\ref{inf-count}) and Sect.~\ref{infra-count-sect}). The X-ray spectrum of the source shows that it has a high absorption, which can be intrinsic absorption by, e.g, the disk boundary layer or stellar wind. The main emission is <5.0~keV and  the low statistics do not allow to significantly verify the hardness of the source and the power-law index is between 1.7--3.4. Its luminosity is  $\sim$1$\times10^{34}$~erg\,s$^{-1}$. Such high luminosity together with highly absorbed emission makes the source a candidate for a $\delta$-symbiotic star. Its X-ray emission would come from the boundary layer of a white dwarf as the acretor. It also can be a $\gamma$-type symbiotic star with a neutron star as the acretor. The nature of the source is not obvious but it is a symbiotic star candidate in  Sculptor.

{\bf Source No.\,126:} The counterpart of the source is classified as a member of Sculptor dSph \citep{2009AJ....137.3100W, 2011MNRAS.414.3492M} and the optical colour magnitude diagram shows that it is on the RGB (see Fog.~\ref{opt-counterpart}).  The source has no 2MASS counterpart and the criteria of \citet{2019MNRAS.483.5077A} can not be checked for this source. The source was only observed in OBS1. As the HRs of the source show, it was only significant in  HR3 and HR4. The spectrum can be described by a power-law model with a photon indexes of 2--3. The lack of significant HR1 and HR2 can be related to the high absorption in the low energies. Since the source was not detected in the second observation it is most probably a variable source and with a luminosity of $\sim1\times10^{34}$~erg\,s$^{-1}$, it might be in the flaring state. The source can be considered as a $\delta$-type symbiotic star candidate. 

{\bf Source No.\,127:} The optical counterpart as  shown in the colour magnitude diagram (see Fig.~\ref{opt-counterpart}) seems to be on the horizontal branch of the Sculptor dSph. \citet{2016MNRAS.462.4349M} classified it as a variable star in the Sculptor dSph. The spectrum of the source shows that the source has a soft component, fit with ionised plasma model (\texttt{apec}), and a power law hard emission. The luminosity of the source is $>10^{34}$~erg\,s$^{-1}$. Assuming that the optical counterpart of the source is still on the main sequence of Sculptor, such high luminosity makes it a candidate for QLMXB.

\section{Summary and Conclusions}

With the aim of  identifying  LMXBs and AWDs, especially symbiotic stars, we have analysed the data of \xmm\, observations of the Sculptor dSph. We  carried out X-ray timing analysis to find variability and pulsation of the sources  using the two methods of Lomb-Scargle and Rayleigh's Z$^{2}_{n}$. No significant pulsation or periodicity was detected for the X-ray sources. Checking the long time variability between the data of two observations did not reveal any significant variable source among the candidates of the X-ray sources in Sculptor dSph either.  We  performed  spectral analysis and HR studies to find the spectral model of the X-ray sources. For the bright X-ray sources, which have been candidates to be members of the Sculptor dSph the spectral analysis has been performed. The HR study gave us information about the spectrum of the X-ray sources, which were not bright enough for spectral analysis to be performed, while the detection likelihood of the X-ray sources was high enough. Moreover, the infrared and optical counterparts of the X-ray sources and from other available catalogues of counterparts were used to identify the possible nature of the X-ray sources. Different methods of the colour and/or magnitude criteria as well as the ratio between the X-ray to optical brightness were used. We classified  43~background AGNs/galaxies and 3~Galactic foreground stars in the field of Sculptor dSph. For the first time in the X-ray study of the Sculptor dSph we classify 7 sources as members of the galaxy:  4~symbiotic stars and  3~QLMXB candidates. This result is consistent with the results of the study of Draco dSph with a stellar mass almost half of  Sculptor dSph \citep[][]{2019A&A...627A.128S}, in which 4~symbiotic stars have been identified. The rest of the sources in the field of Sculptor dSph remain unclassified due to the lack of information in X-rays and/or for the multi-wavelength counterparts.
\section*{Acknowledgements}
This  research  was  funded  through  the BMWi/DLR research grant 50OR1907. This study is based on observations obtained with \xmm, an ESA science mission with instruments  and  contributions  directly  funded  by  ESA  Member  States  and  NASA. This  research  has  made  use  of  the  SIMBAD  and  VIZIER  database,  operated at  CDS,  Strasbourg,  France,  and  of  the  NASA/IPAC  Extra galactic  Database\,(NED),  which  is  operated  by  the  Jet  Propulsion  Laboratory,  California  Institute  of  Technology,  under  contract  with  the  National  Aeronautics  and  Space Administration. This publication makes use of data products from the Wide field Infrared Survey Explorer, which is a joint project of the University of California, Los Angeles, and the Jet Propulsion Laboratory/California Institute of Technology, funded by the National Aeronautics and Space Administration. This publication has made use of data products from the Two Micron All Sky Survey, which is a joint project of the University of Massachusetts and the Infrared Processing  and  Analysis  Center,  funded  by  the  National  Aeronautics  and  Space Administration  and  the  National  Science  Foundation.  Funding  for  SDSS  and SDSS-III has been provided by the Alfred P. Sloan Foundation, the Participating  Institutions,  the  National  Science  Foundation,  and  the  US  Department  of Energy Office of Science. The SDSS-III web site ishttp://www.sdss3.org/.SDSS-III  is  managed  by  the  Astrophysical  Research  Consortium  for  the  Participating  Institutions  of  the  SDSS-III  Collaboration  including  the  University of  Arizona,  the  Brazilian  Participation  Group,  Brookhaven  National  Laboratory, University of Cambridge, University of Florida, the French Participation Group, the German Participation Group, the Instituto de Astrofisica de Canarias, the Michigan State/Notre Dame/JINA Participation Group, Johns Hopkins University,  Lawrence  Berkeley  National  Laboratory,  Max  Planck  Institute  for Astrophysics, New Mexico State University, New York University, Ohio State University, Pennsylvania State University, University of Portsmouth, Princeton University, the Spanish Participation Group, University of Tokyo, University of Utah, Vanderbilt University, University of Virginia, University of Washington, and Yale University. This research has made use of SAO Image DS9, developed by Smithsonian Astrophysical Observatory.

\section*{Data Availability}

 This work is based on the three observations of \xmm\, already 
accessible at https://heasarc.gsfc.nasa.gov/docs/archive.html. The 
details of data are available in the appendix tables (see Tables \ref{catalogue-x-ray}, \ref{opt-count-table}, and \ref{inf-count}. The details of data analysis are explained in the paper, and more details of programming underlying this article will be shared
on reasonable request to the corresponding author.



\bibliographystyle{mnras}
\bibliography{example} 




\appendix
\clearpage
\onecolumn


\onecolumn
\begin{landscape}
\raggedright
\section{Source catalogue}

\setlength{\tabcolsep}{1.1mm}
\small{
\begin{longtable}{cccccccccccl}
    \caption{X-ray sources in the field of Sculptor dSph. \label{catalogue-x-ray}}\tabularnewline
\hline
NO   &RA              &DEC              &r1$\sigma$                       & \multicolumn{2}{c}{Flux (0.2--4.5~keV)}  &\multicolumn{4}{l}{Hardness ratio} &Var&Classification$^{\dagger}$\\
                          &  (J2000)       &  (J2000)        & ($\arcsec$)                     & \multicolumn{2}{c}{($10^{-14}$ erg\,s$^{-1}$\,cm$^{-2}$)}&&&&&\\
                           &                &                 &                                 & OBS1&          OBS2&       HR1&HR2&HR3&HR4&&\\
\hline
\endfirsthead
\multicolumn {12}{r}{Continued from previous page}\\
\hline 
 NO  &RA              &DEC              &r1$\sigma$                       & \multicolumn{2}{c}{Flux (0.2--4.5~keV)} &\multicolumn{4}{l}{Hardness ratio} &Var& Classification$^{\dagger}$\\
                     &  (J2000)       &  (J2000)        & ($\arcsec$)                     & \multicolumn{2}{c}{($10^{-14}$ erg\,s$^{-1}$\,cm$^{-2}$)}&&&&&\\
                     &                &                 &                                 & OBS1&          OBS2&    HR1&HR2&HR3&HR4&&\\
\hline
\endhead
\hline
\multicolumn {12}{r}{Continued on the next page}\\

\endfoot

\hline
\endlastfoot

1&         00 58 54.28 &     -33 44 02.1  &            2.06  &                1.75$\pm$                0.21 &                -- &               -0.06$\pm$                0.16 &                -- &                -- &                -- &                -- &  AGN    \\
      2&         00 58 58.88 &     -33 42 31.2  &            2.47  &                2.64$\pm$                0.27 &                -- &                -- &                -- &                -- &                -- &                -- &  AGN    \\
      3&         00 58 58.99 &     -33 42 00.2  &            1.98  &                4.91$\pm$                0.32 &                -- &                -- &                0.20$\pm$                0.14 &               -0.24$\pm$                0.17 &                -- &                -- &  AGN    \\
      4&         00 59 03.25 &     -33 40 20.0  &            2.04  &                1.07$\pm$                0.13 &                1.85$\pm$                0.17 &               -0.12$\pm$                0.21 &               -0.26$\pm$                0.16 &                -- &                -- &                1.73$\pm$                0.22 &      \\
      5&         00 59 06.11 &     -33 49 35.7  &            1.36  &                3.24$\pm$                0.28 &                -- &               -0.03$\pm$                0.19 &               -0.10$\pm$                0.19 &               -0.01$\pm$                0.23 &                -- &                -- &      \\
      6&         00 59 08.95 &     -33 36 42.5  &            2.10  &                1.25$\pm$                0.45 &                0.64$\pm$                0.18 &                -- &                -- &                -- &                -- &                1.95$\pm$                0.64 &      \\
\hline
  
\end{longtable}}
\end{landscape}

\section{Optical and infrared magnitudes of counterparts of  X-ray sources.}
\label{oni-tables}

\begin{longtable}{lccccc}
    \caption{ Optical magnitudes of counterparts of X-ray sources in different energy filters of the SDSS9 survey.\label{opt-count-table}}\tabularnewline
\hline
Src-No&  $g$ mag&  $r$   mag&  $i$ mag&  $z$ mag&  $y$ mag\\
\hline
\endfirsthead
\multicolumn {6}{r}{Continued from previous page}\\
\hline 
 Src-No&$g$ mag&$r$ mag&$i$ mag&$z$ mag&$y$ mag\\
\hline
\endhead
\hline
\multicolumn {6}{r}{Continued on the next page}\\

\endfoot

\hline
\endlastfoot

 2&               20.37$\pm$                0.02 &               19.21$\pm$                0.01 &               18.81$\pm$                0.01 &               18.50$\pm$                0.01 &               18.46$\pm$                0.03     \\
      4&               22.68$\pm$                0.08 &               22.07$\pm$                0.07 &               21.95$\pm$                0.09 &               21.93$\pm$                0.12 &               22.09$\pm$                0.30     \\
      5&               23.27$\pm$                0.07 &               22.79$\pm$                0.05 &               22.32$\pm$                0.07 &               22.09$\pm$                0.10 &               22.01$\pm$                0.24     \\
      6&               22.27$\pm$                0.05 &               21.13$\pm$                0.03 &               20.09$\pm$                0.02 &               19.64$\pm$                0.02 &               19.45$\pm$                0.05     \\
      7&               23.40$\pm$                0.11 &               23.13$\pm$                0.11 &               23.70$\pm$                0.35 &               23.05$\pm$                0.38 &               23.32$\pm$                1.18     \\
      8&               23.04$\pm$                0.06 &               22.67$\pm$                0.07 &               22.47$\pm$                0.09 &               22.53$\pm$                0.18 &               23.90$\pm$                1.56     \\
     10&               24.06$\pm$                0.13 &               23.98$\pm$                0.17 &               23.94$\pm$                0.30 &               24.02$\pm$                0.61 &               23.40$\pm$                0.87     \\
     11&               21.69$\pm$                0.03 &               20.57$\pm$                0.02 &               19.88$\pm$                0.02 &               19.55$\pm$                0.02 &               19.44$\pm$                0.05     \\
\end{longtable}

\begin{longtable}{lccccccc}
    \caption{Infrared magnitudes of counterparts of X-ray sources in different energy filters of 2MASS and WISE surveys.\label{inf-count}}\tabularnewline
\hline
Src-No&$W1$ mag&$W2$ mag&$W3$ mag&$W4$ mag&$J$ mag&$H$ mag&$K$ mag\\
\hline
\endfirsthead
\multicolumn {8}{r}{Continued from previous page}\\
\hline 
Src-No&$W1$ mag&$W2$ mag&$W3$ mag&$W4$ mag&$J$ mag&$H$ mag&$K$ mag\\
\hline
\endhead
\hline
\multicolumn {8}{r}{Continued on the next page}\\

\endfoot

\hline
\endlastfoot
       1&               17.01$\pm$                0.11 &               15.78$\pm$                0.13 &               12.41$\pm$                0.43 &                9.15$\pm$                0.00 &                -- &                -- &                --     \\
      2&               15.26$\pm$                0.04 &               14.83$\pm$                0.07 &               11.88$\pm$                0.26 &                8.44$\pm$                0.00 &                -- &                -- &                --     \\
      3&               16.88$\pm$                0.09 &               15.78$\pm$                0.12 &               12.55$^{\ast}$  &                8.98$^{\ast}$ &                -- &                -- &                --     \\
      4&               17.34$\pm$                0.13 &               17.39$\pm$                0.52 &               12.74$^{\ast}$  &                9.02$^{\ast}$ &                -- &                -- &                --     \\
      6&               15.40$\pm$                0.04 &               15.24$\pm$                0.08 &               12.35$^{\ast}$  &                9.07$^{\ast}$ &                -- &                -- &                --     \\
      7&               16.75$\pm$                0.09 &               17.26$\pm$                0.50 &               12.34$^{\ast}$  &                9.11$^{\ast}$ &                -- &                -- &                --     \\
      8&               16.69$\pm$                0.08 &               16.47$\pm$                0.22 &               12.31$^{\ast}$  &                9.00$^{\ast}$ &                -- &                -- &                --     \\
      9&               17.27$\pm$                0.13 &               16.18$\pm$                0.17 &               12.26$^{\ast}$  &                8.82$^{\ast}$ &                -- &                -- &                --     \\
\end{longtable}

\section{Image of optical counterparts}
%
\begin{figure}
\caption{SDSS7 optical image of the counterpart of the X-ray sources, which have been discussed in Sect.\ref{diss}. The red circles show the $3\sigma$ error of X-ray source positions. The size of all images is $32.\arcsec\times32.\arcsec$. In the image\,(a) of Src-No.13 a scale bar of 5.0$\arcsec$ is shown. \label{opt-image}}
\centering
 \subfloat[Src-No.\,13  ]{\includegraphics[clip, trim={0.0cm 2.cm 0.cm 0.0cm},width=0.26\textwidth]{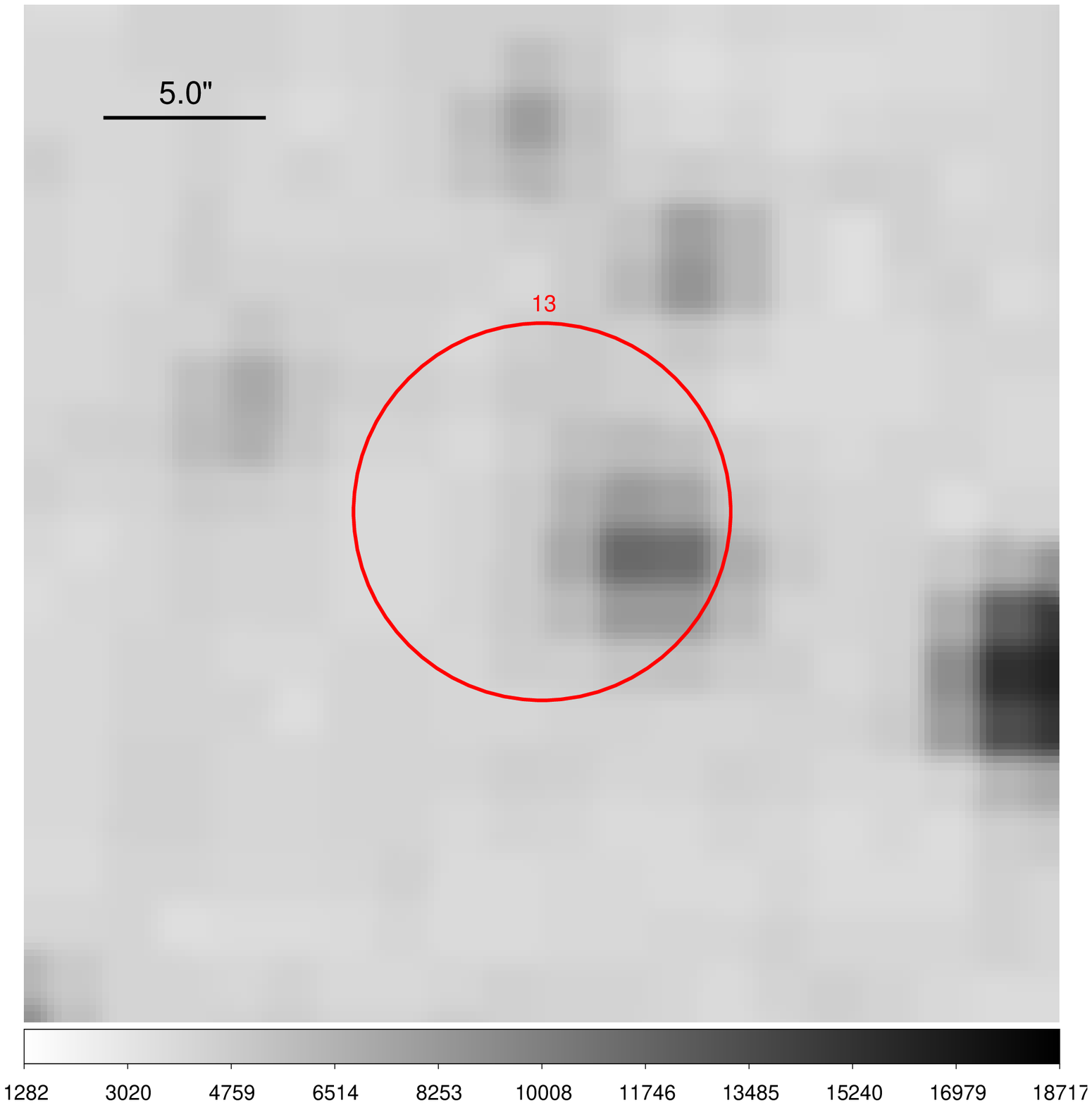}}  
  \subfloat[Src-No.\,22  ]{\includegraphics[clip, trim={0.0cm 2.cm 0.cm 0.0cm},width=0.26\textwidth]{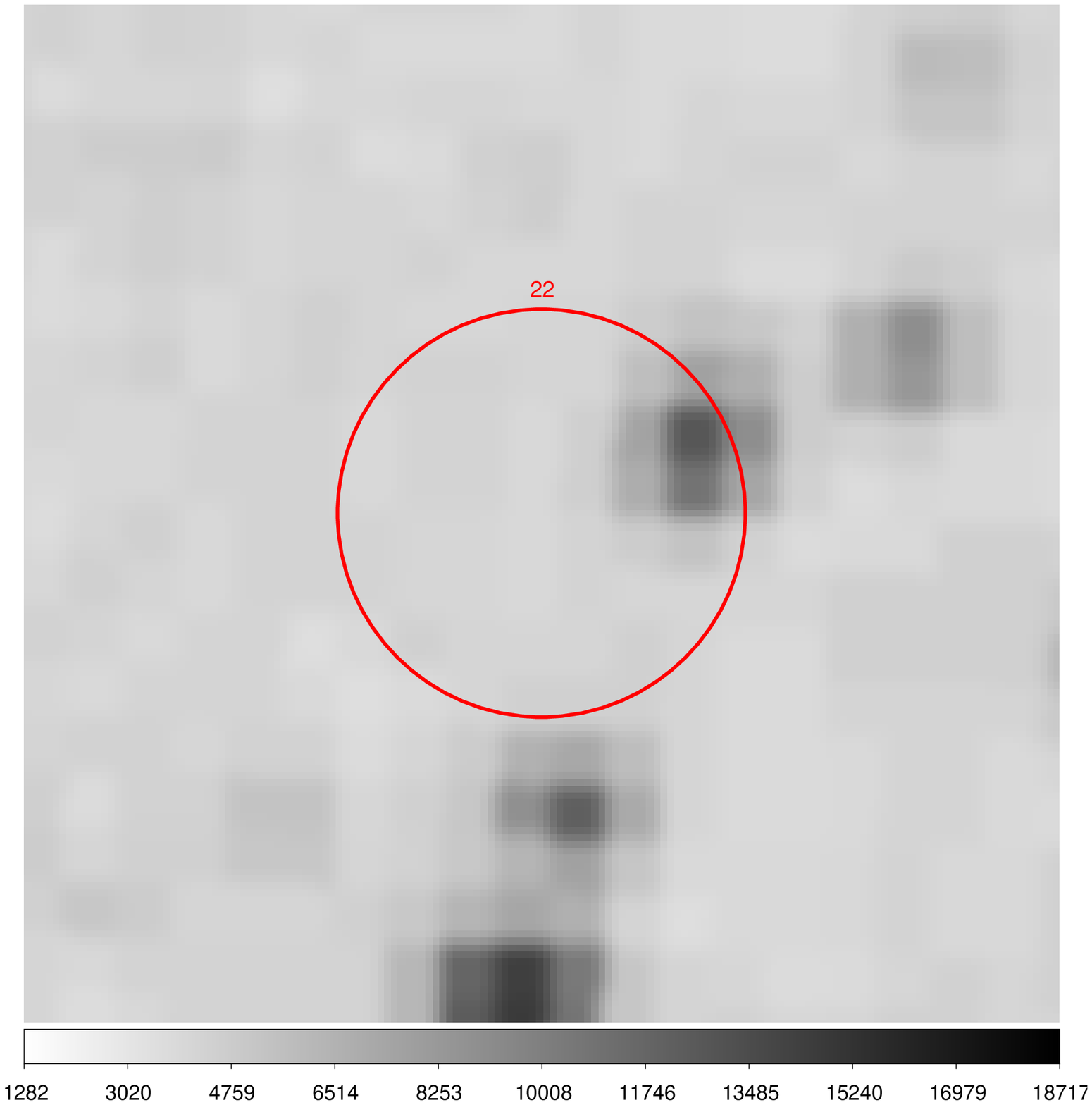}}
  \subfloat[Src-No.\,23  ]{\includegraphics[clip, trim={0.0cm 2.cm 0.cm 0.0cm},width=0.26\textwidth]{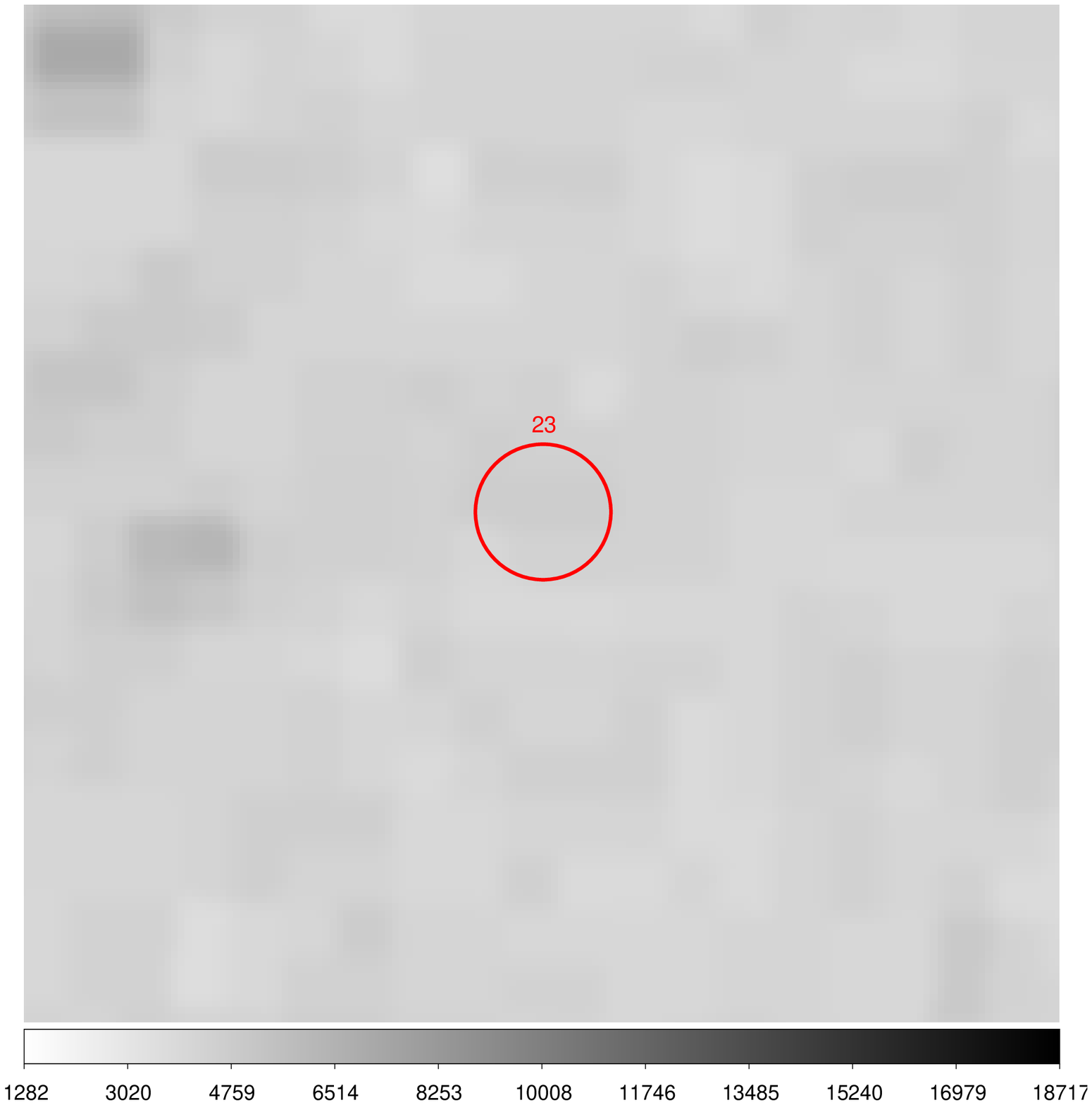}} \\ 
  \subfloat[Src-No.\,27  ]{\includegraphics[clip, trim={0.0cm 2.cm 0.cm 0.0cm},width=0.26\textwidth]{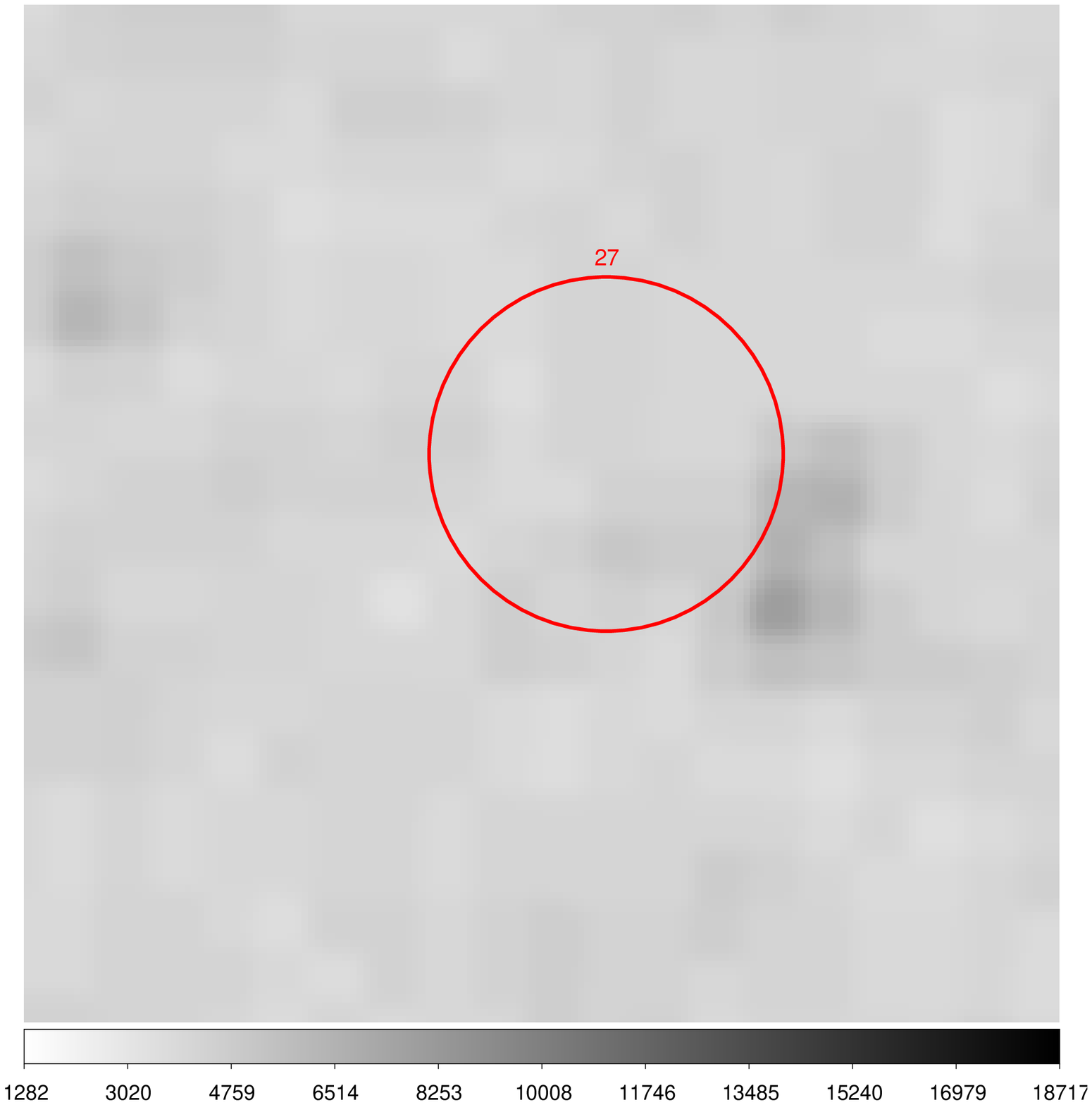}}  
  \subfloat[Src-No.\,52  ]{\includegraphics[clip, trim={0.0cm 2.cm 0.cm 0.0cm},width=0.26\textwidth]{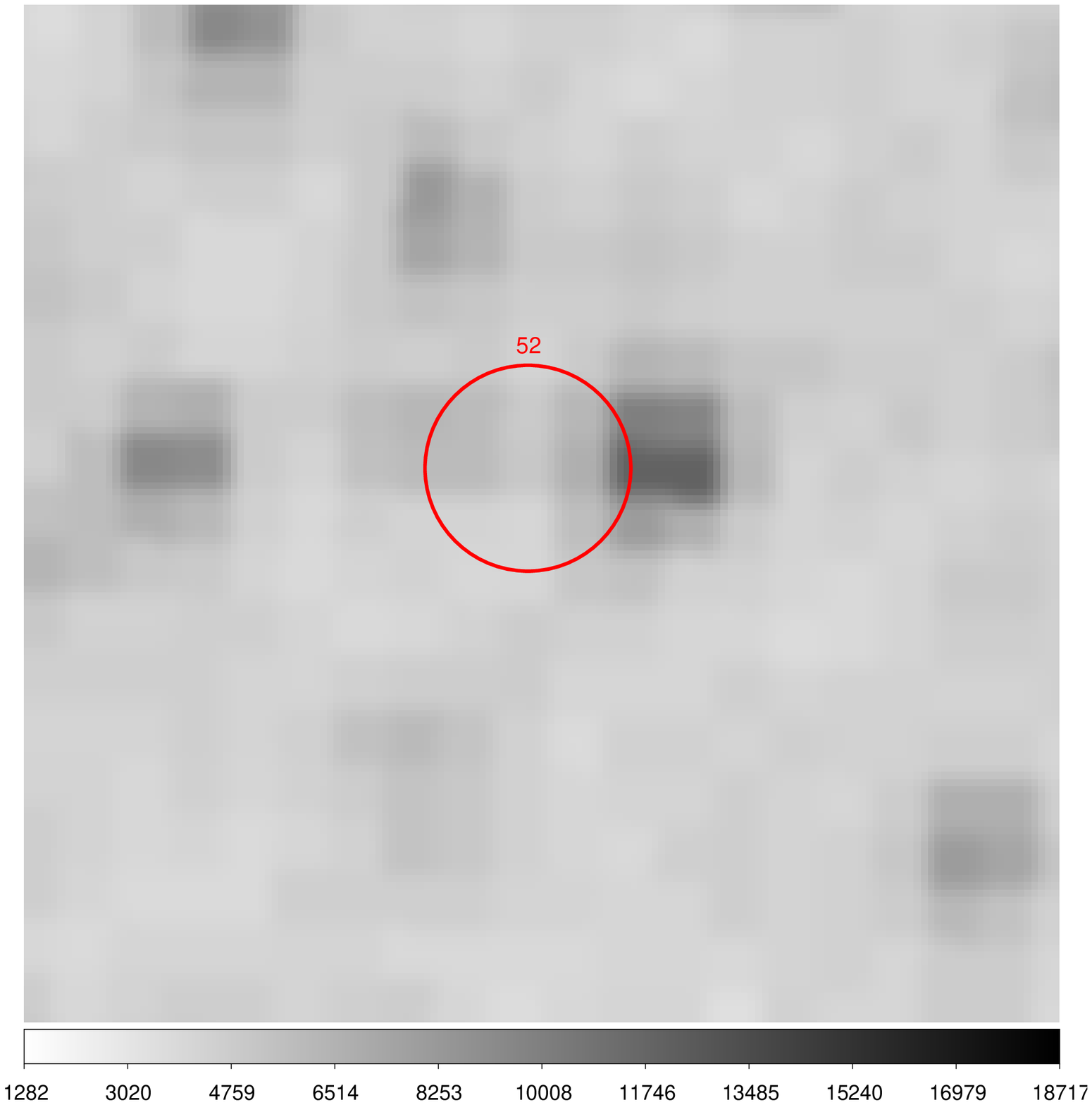}}
  \subfloat[Src-No.\,58  ]{\includegraphics[clip, trim={0.0cm 2.cm 0.cm 0.0cm},width=0.26\textwidth]{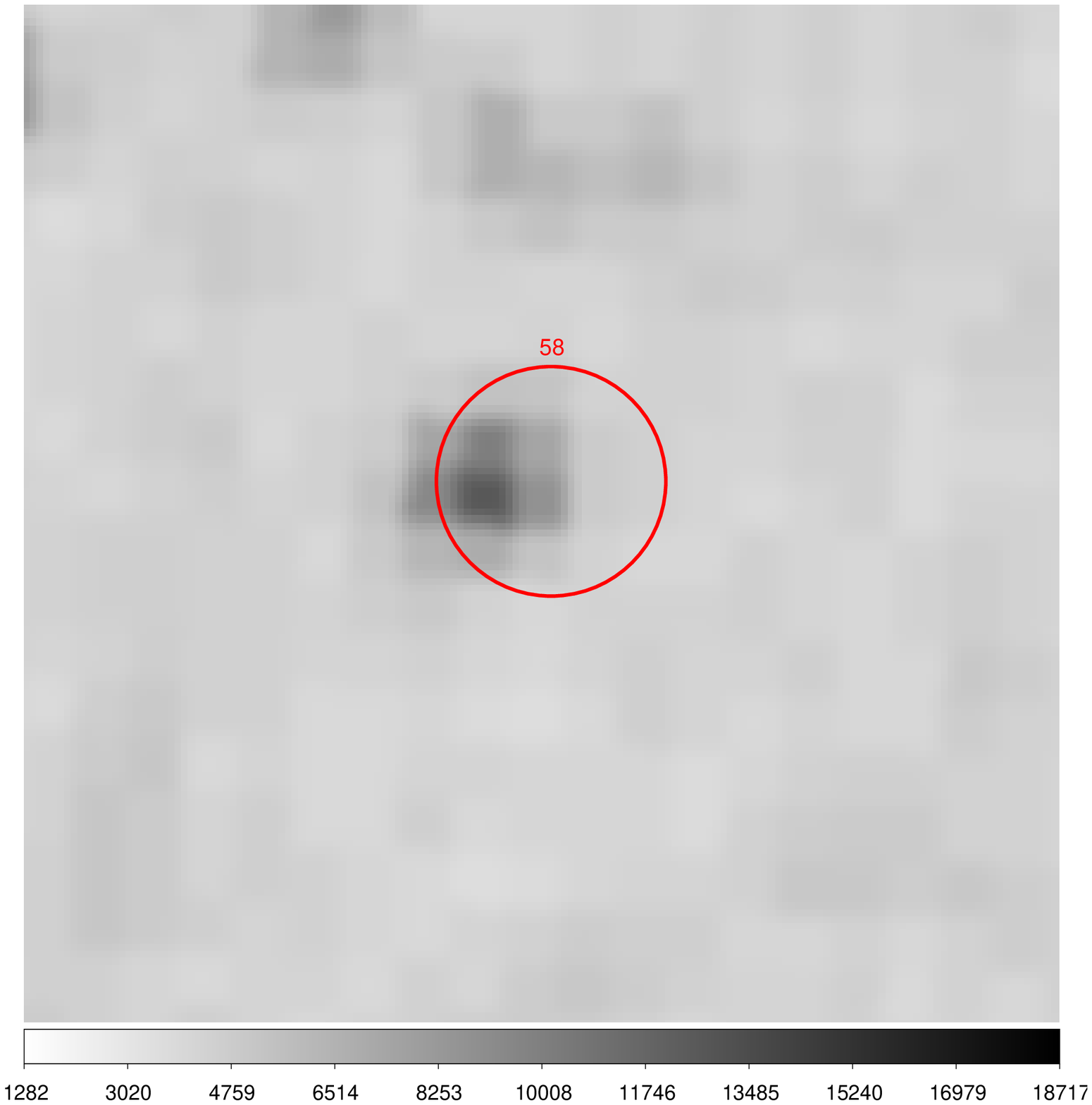}}  \\
  \subfloat[Src-No.\,61  ]{\includegraphics[clip, trim={0.0cm 2.cm 0.cm 0.0cm},width=0.26\textwidth]{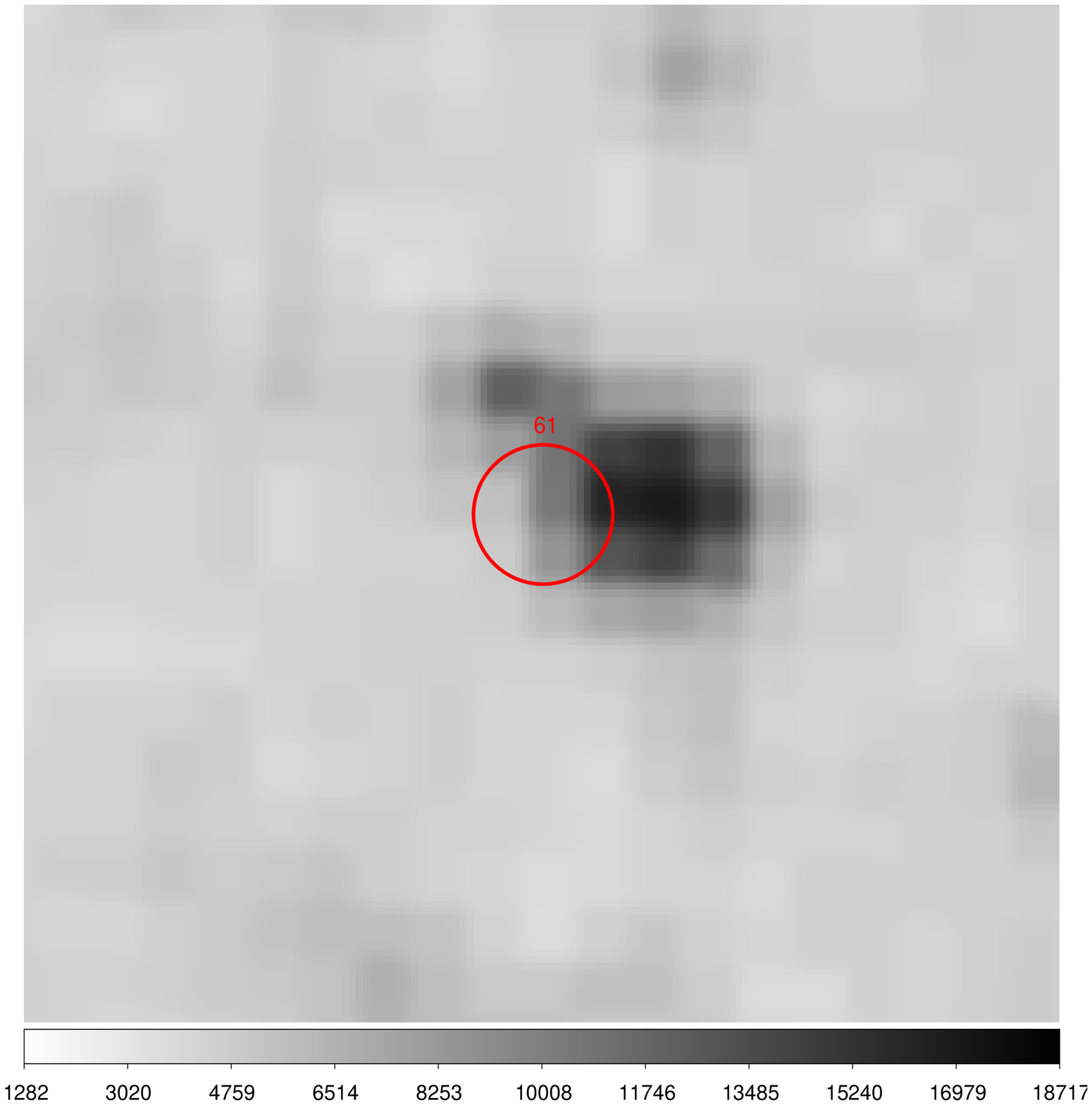}}  
  \subfloat[Src-No.\,62  ]{\includegraphics[clip, trim={0.0cm 2.cm 0.cm 0.0cm},width=0.26\textwidth]{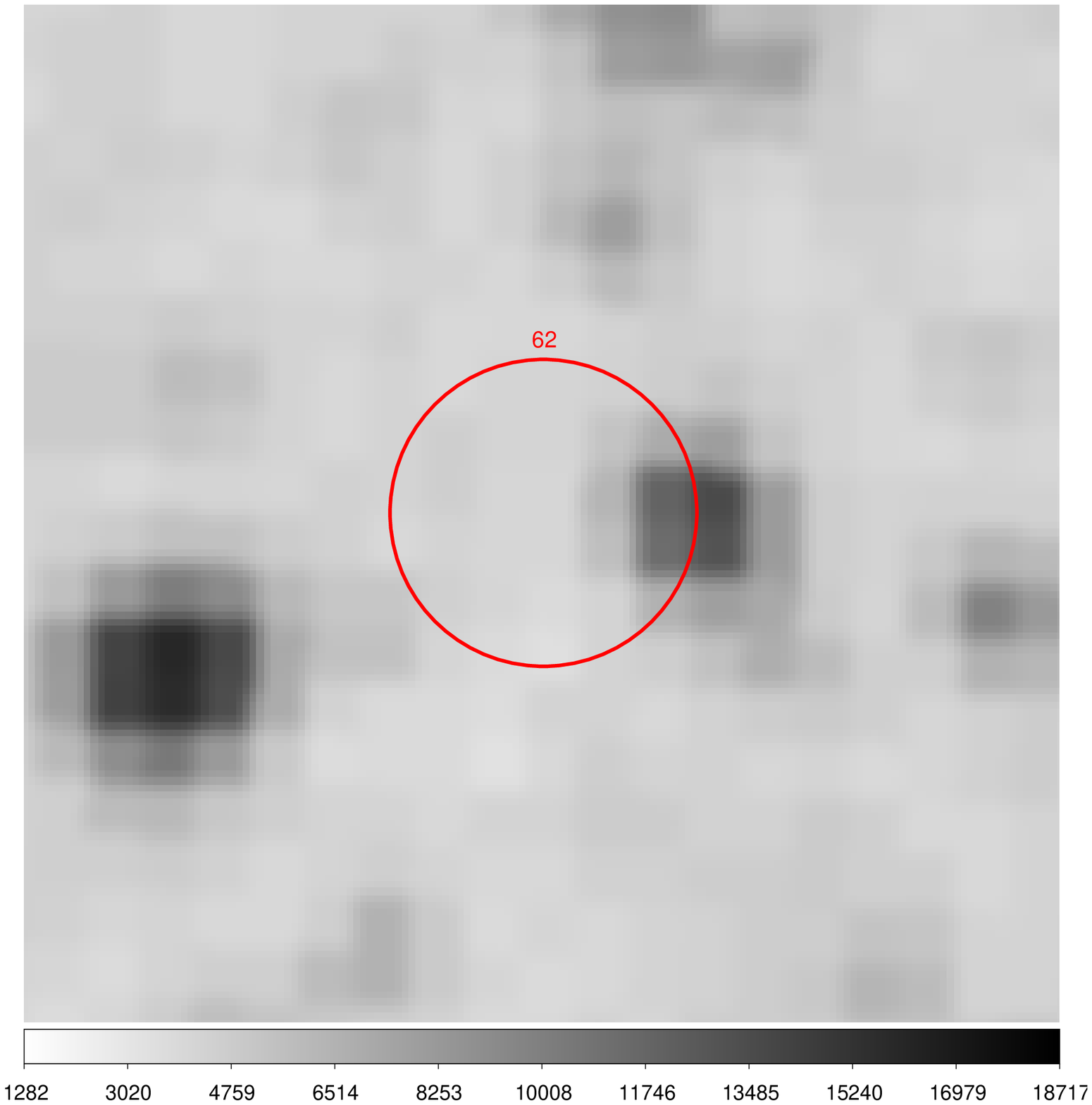}}
  \subfloat[Src-No.\,68  ]{\includegraphics[clip, trim={0.0cm 2.cm 0.cm 0.0cm},width=0.26\textwidth]{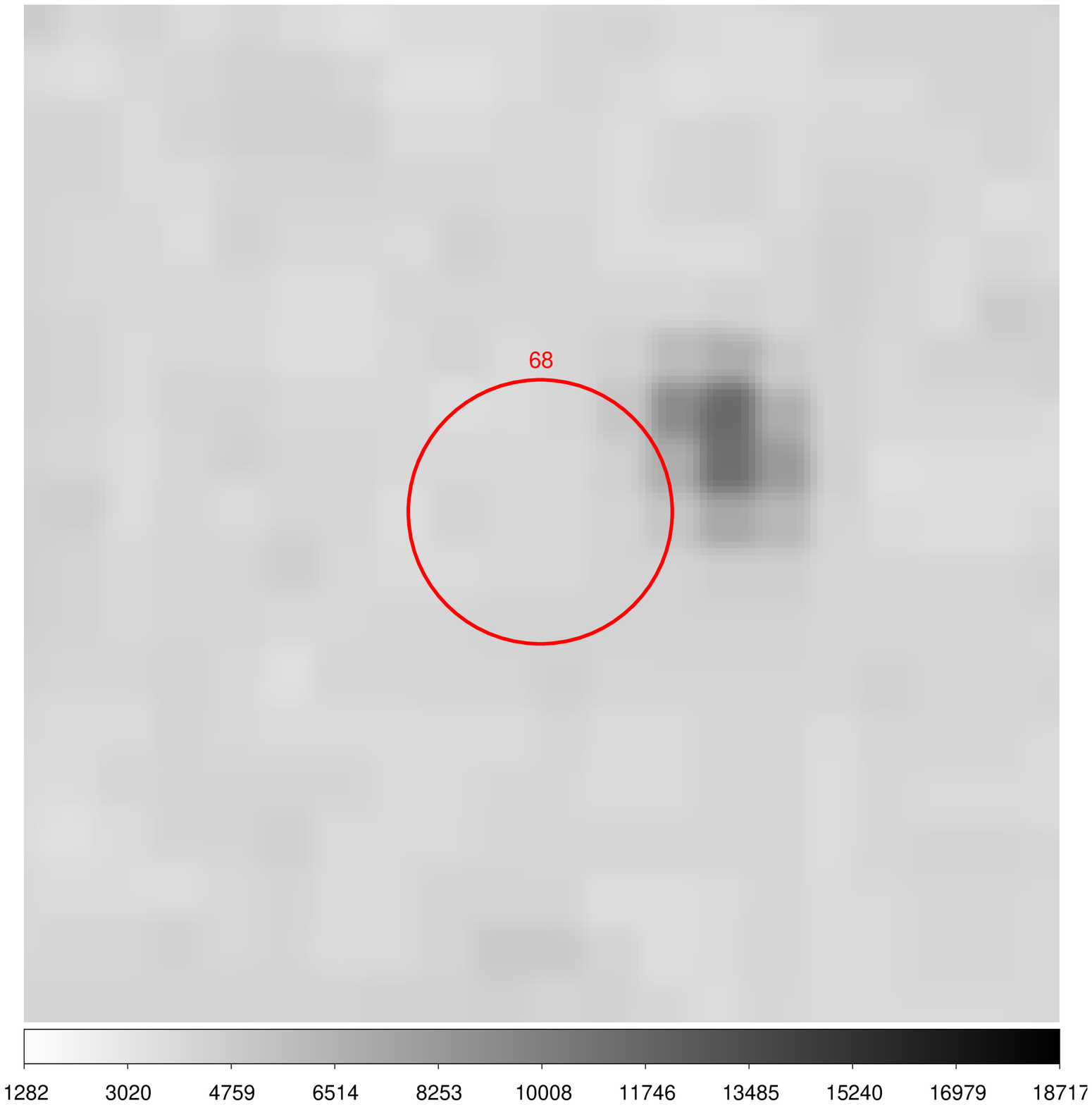}}  \\
 \subfloat[Src-No.\,69  ]{\includegraphics[clip, trim={0.0cm 2.cm 0.cm 0.0cm},width=0.26\textwidth]{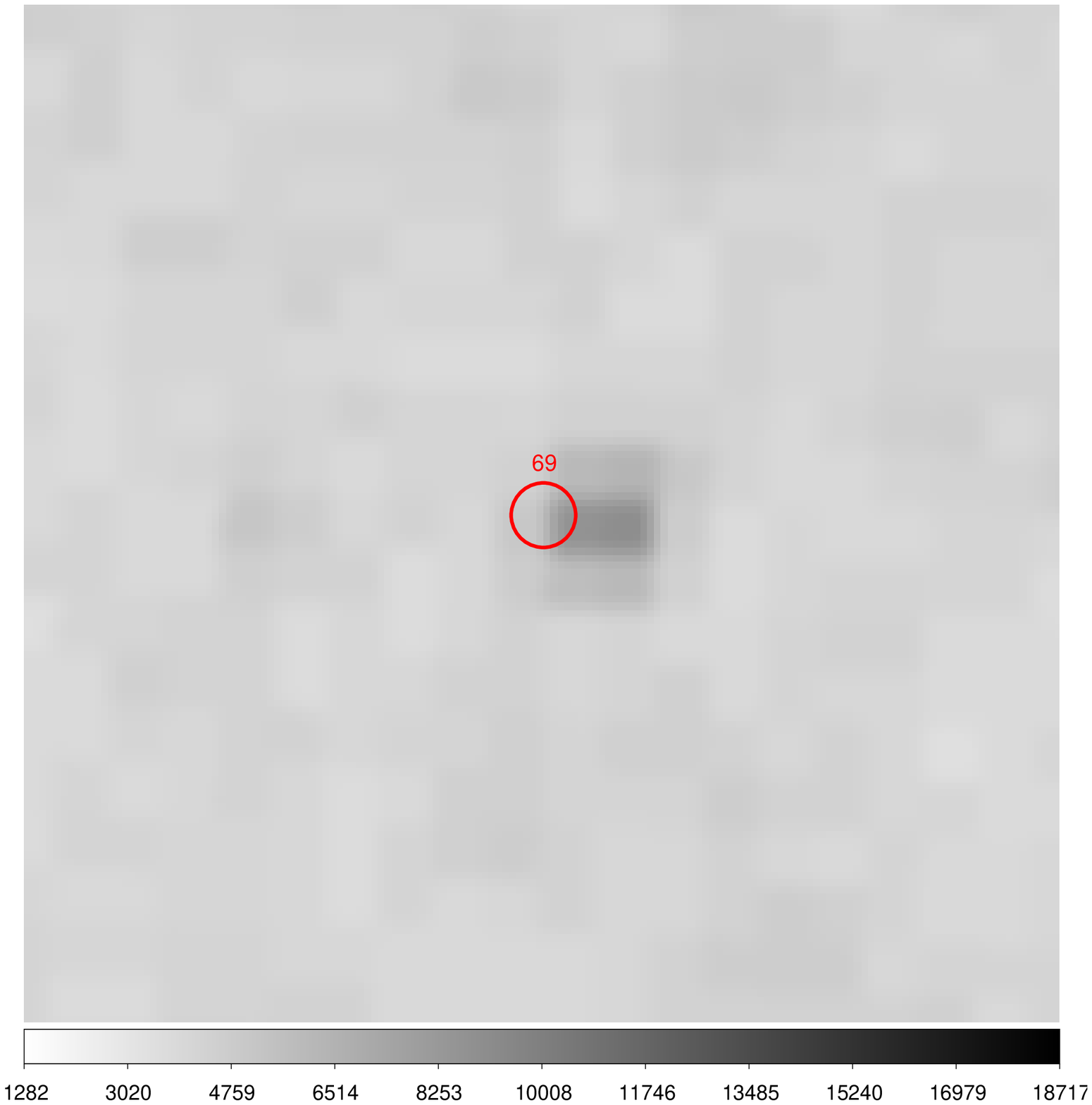}}  
 \subfloat[Src-No.\,73 ]{\includegraphics[clip, trim={0.0cm 2.cm 0.cm 0.0cm},width=0.26\textwidth]{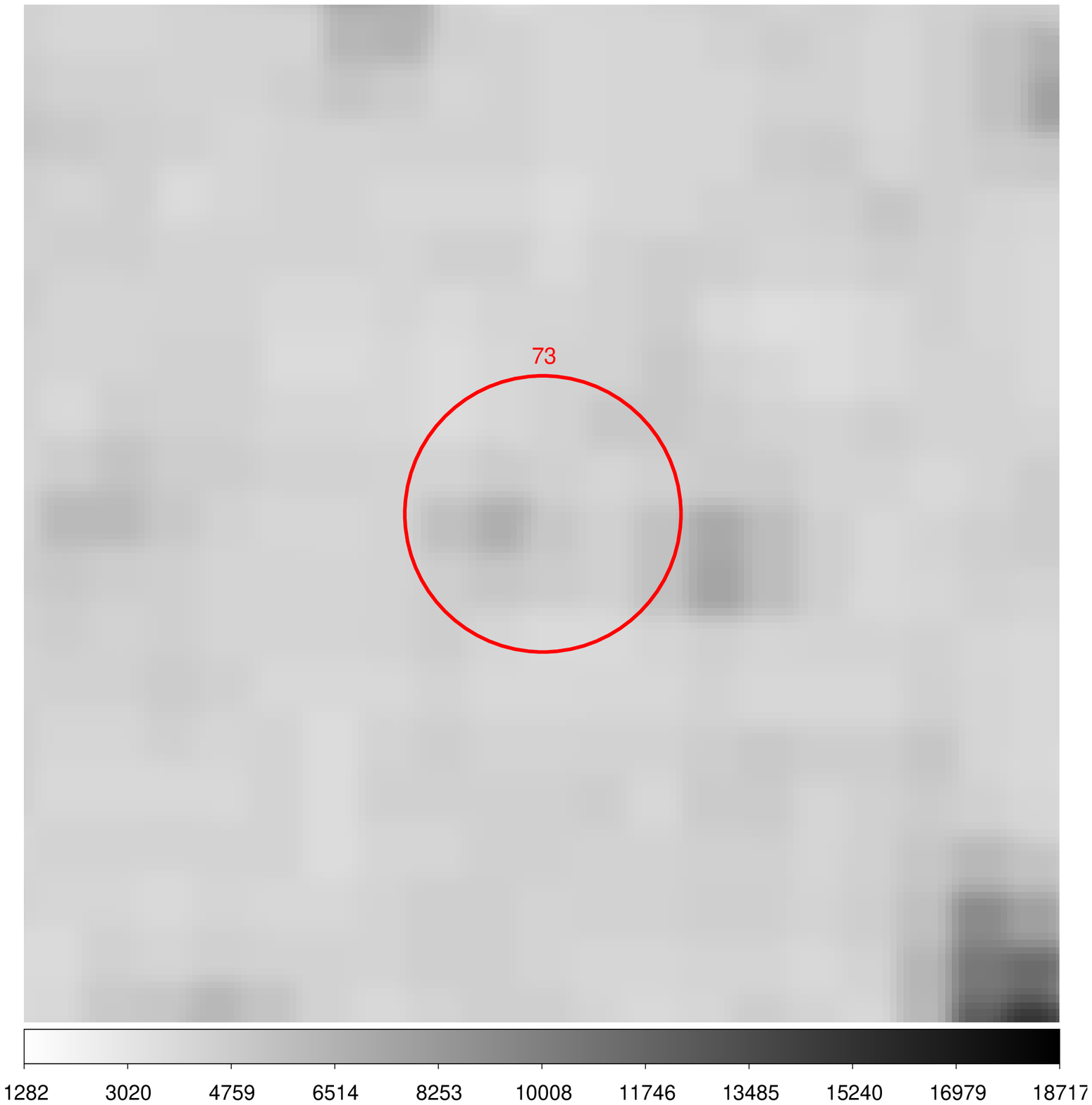}}
 \subfloat[Src-No.\,74  ]{\includegraphics[clip, trim={0.0cm 2.cm 0.cm 0.0cm},width=0.26\textwidth]{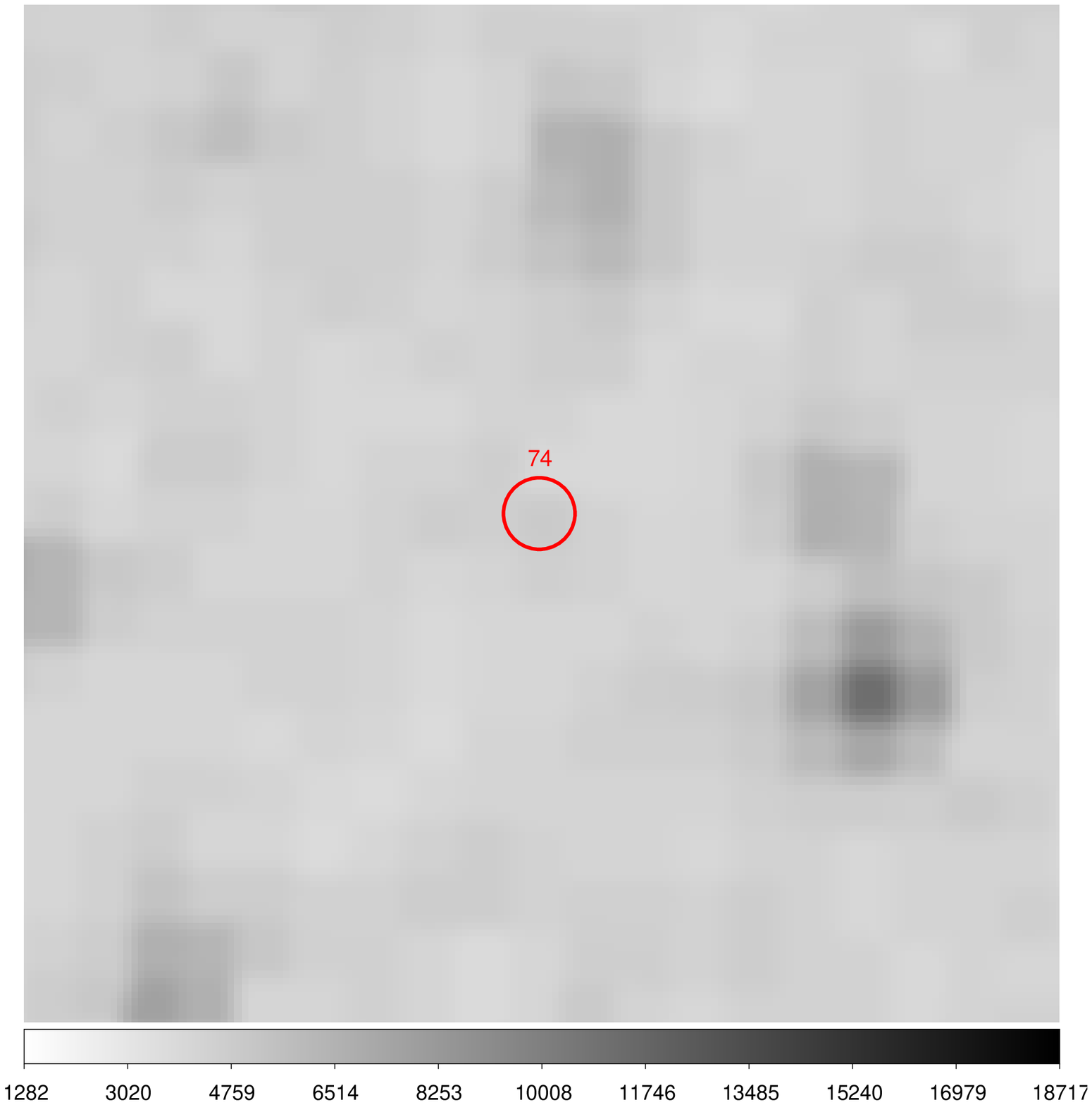}}  \\
\end{figure}
\pagebreak
\onecolumn
\clearpage
\begin{figure}
\centering
\vspace{-0.5cm}
\subfloat[Src-No.\,83  ]{\includegraphics[clip, trim={0.0cm 2.cm 0.cm 0.0cm},width=0.26\textwidth]{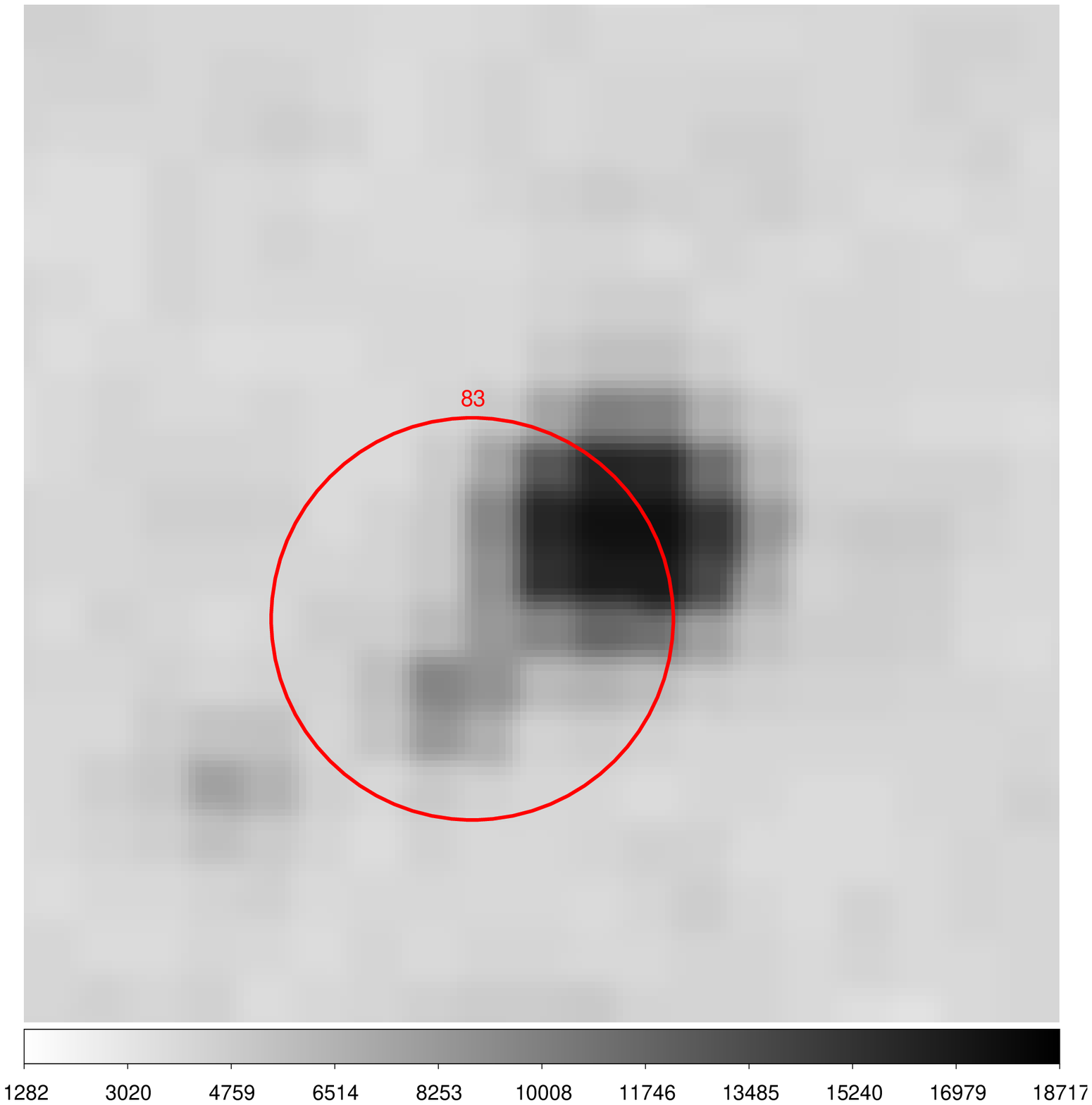}}  
  \subfloat[Src-No.\,88  ]{\includegraphics[clip, trim={0.0cm 2.cm 0.cm 0.0cm},width=0.26\textwidth]{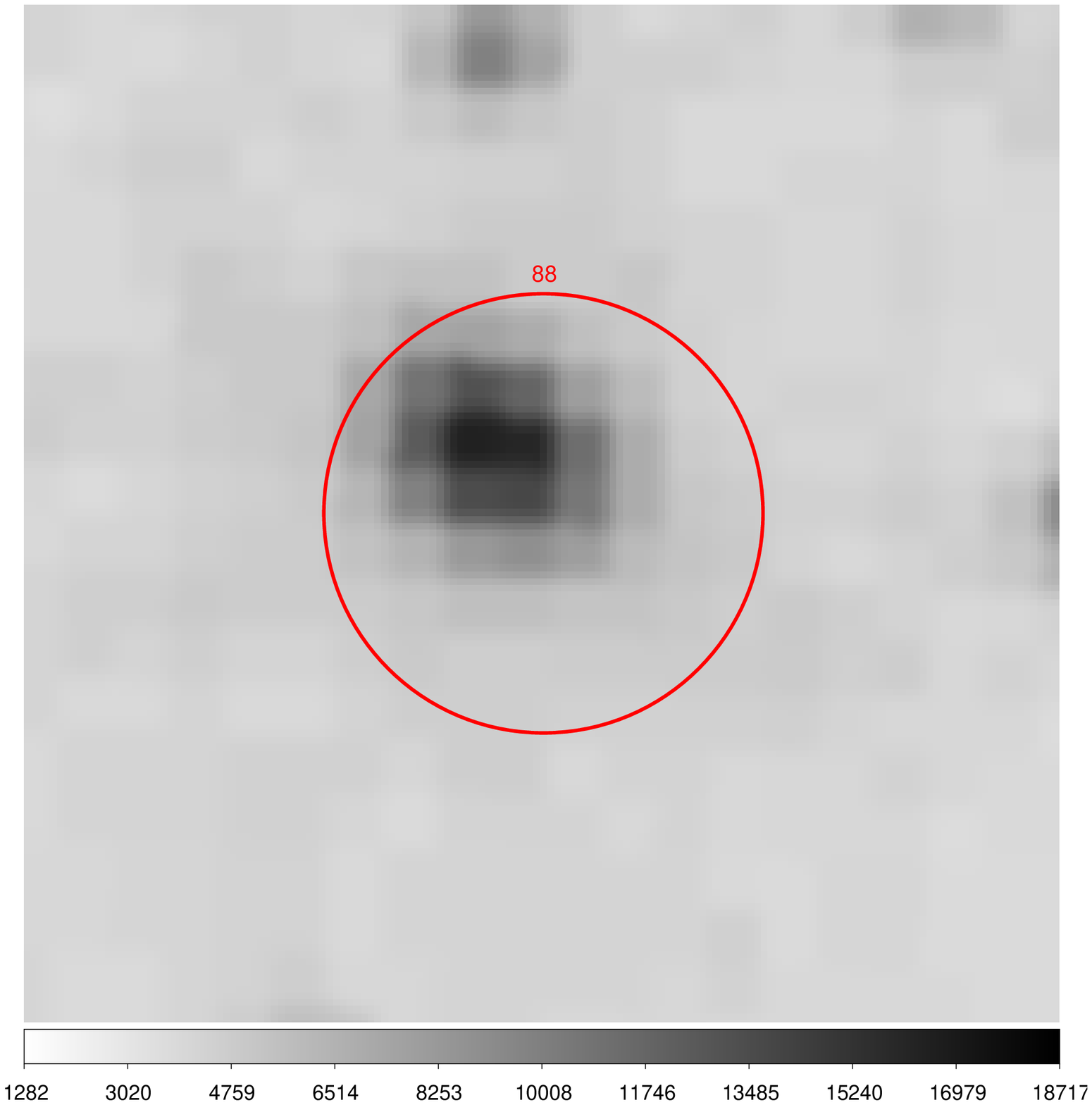}}
  \subfloat[Src-No.\,100  ]{\includegraphics[clip, trim={0.0cm 2.cm 0.cm 0.0cm},width=0.26\textwidth]{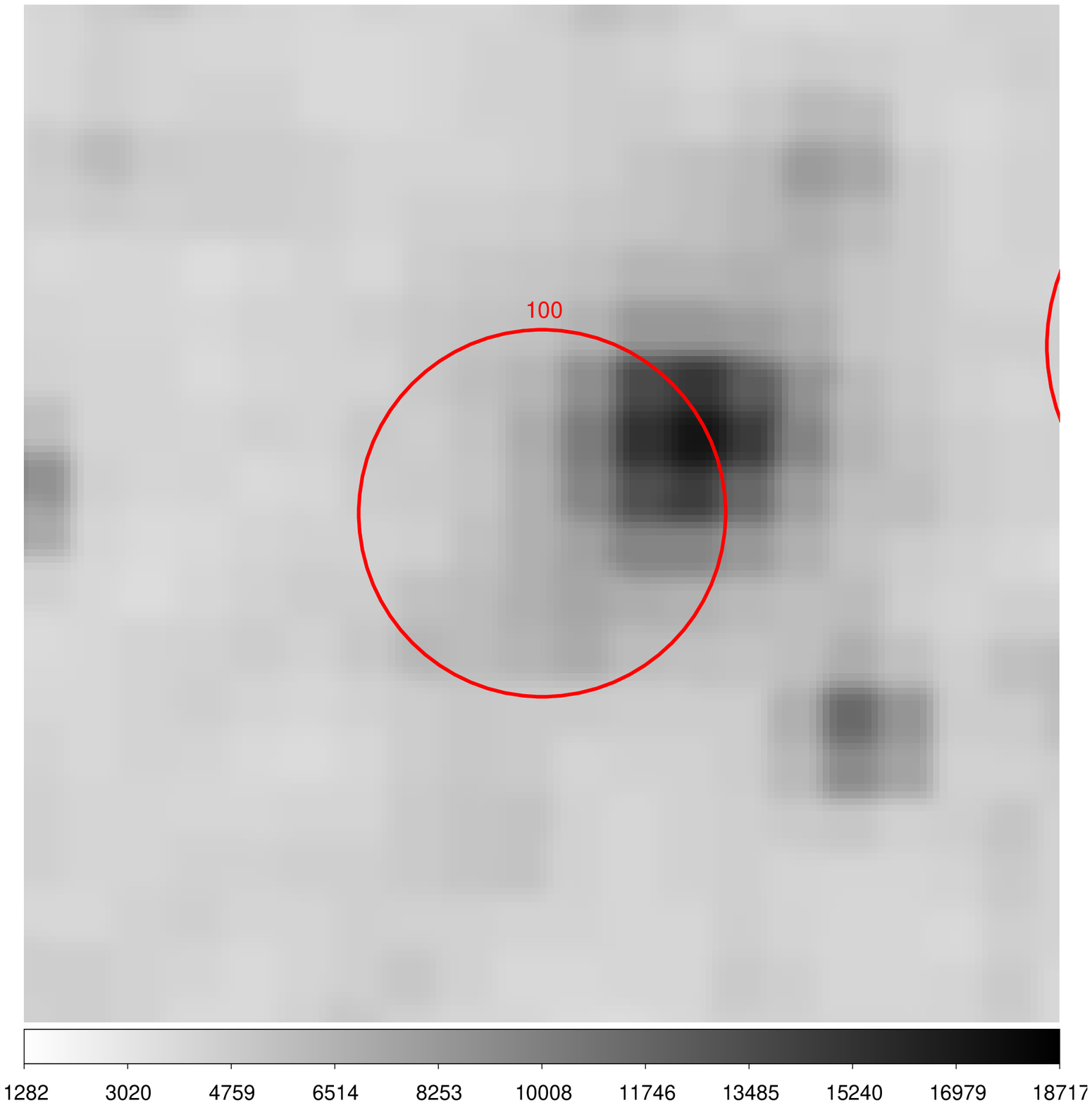}}  \\
  \subfloat[Src-No.\,101  ]{\includegraphics[clip, trim={0.0cm 2.cm 0.cm 0.0cm},width=0.26\textwidth]{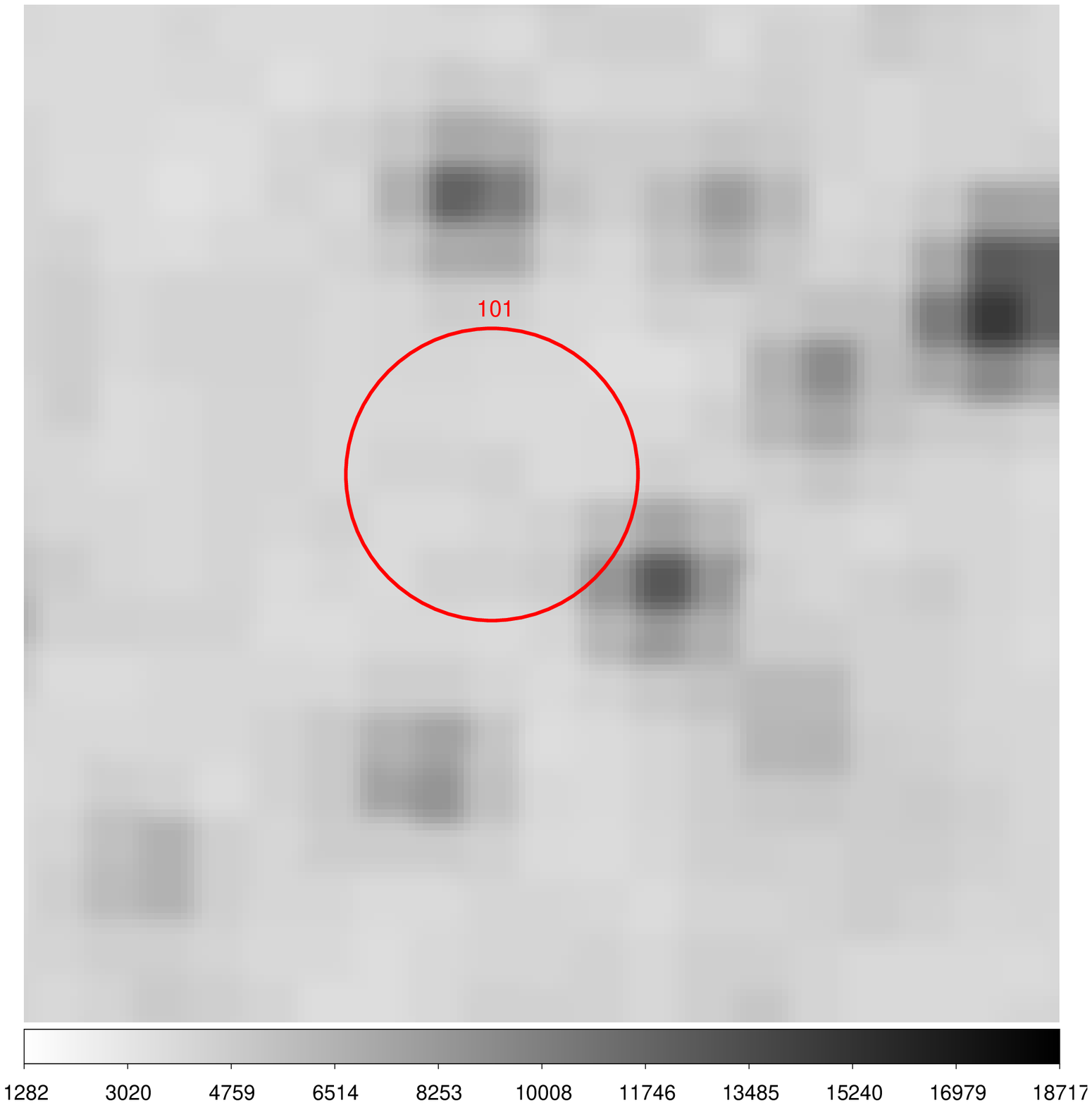}}  
  \subfloat[Src-No.\,109  ]{\includegraphics[clip, trim={0.0cm 2.cm 0.cm 0.0cm},width=0.26\textwidth]{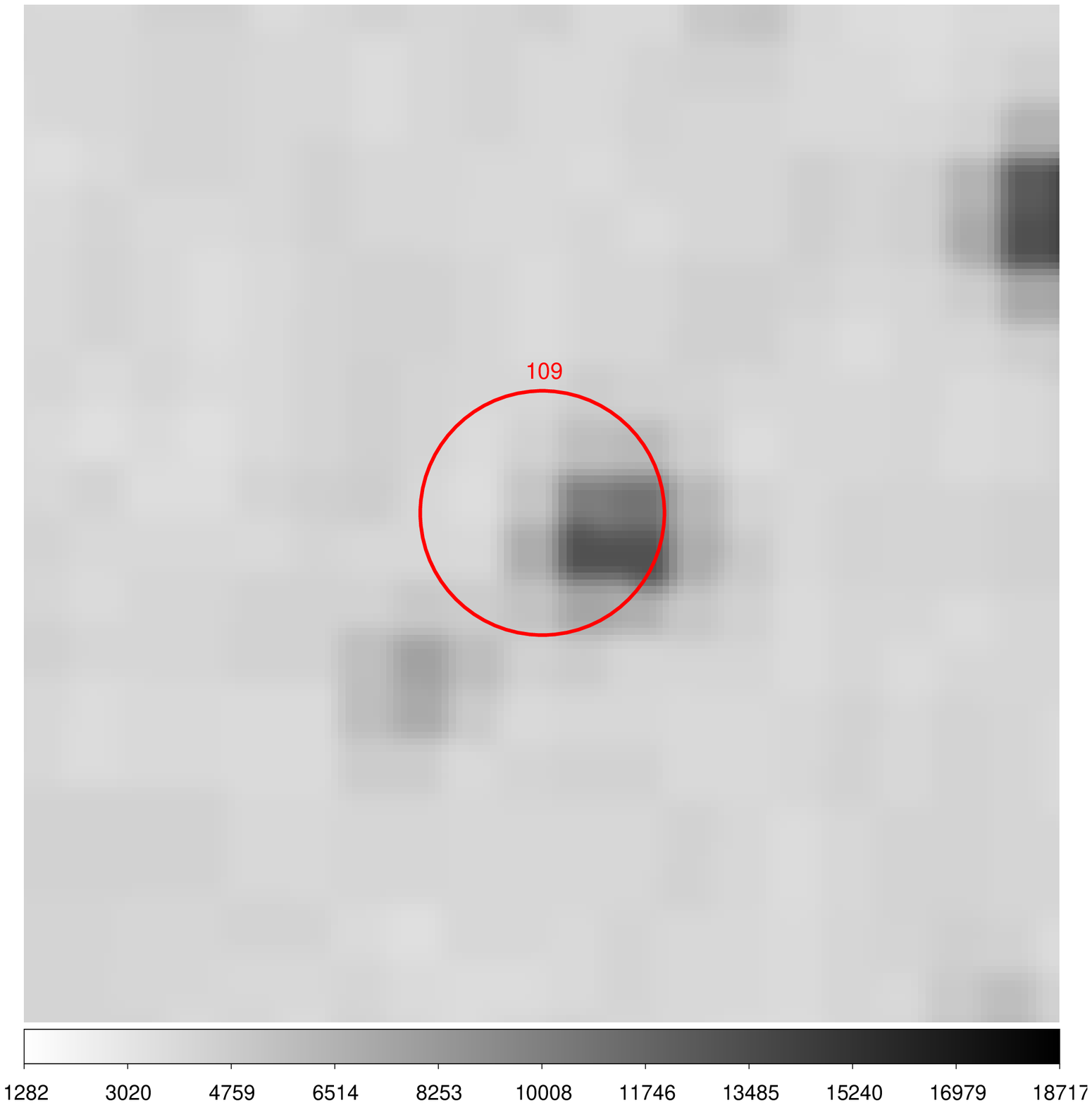}}
  \subfloat[Src-No.\,126  ]{\includegraphics[clip, trim={0.0cm 2.cm 0.cm 0.0cm},width=0.26\textwidth]{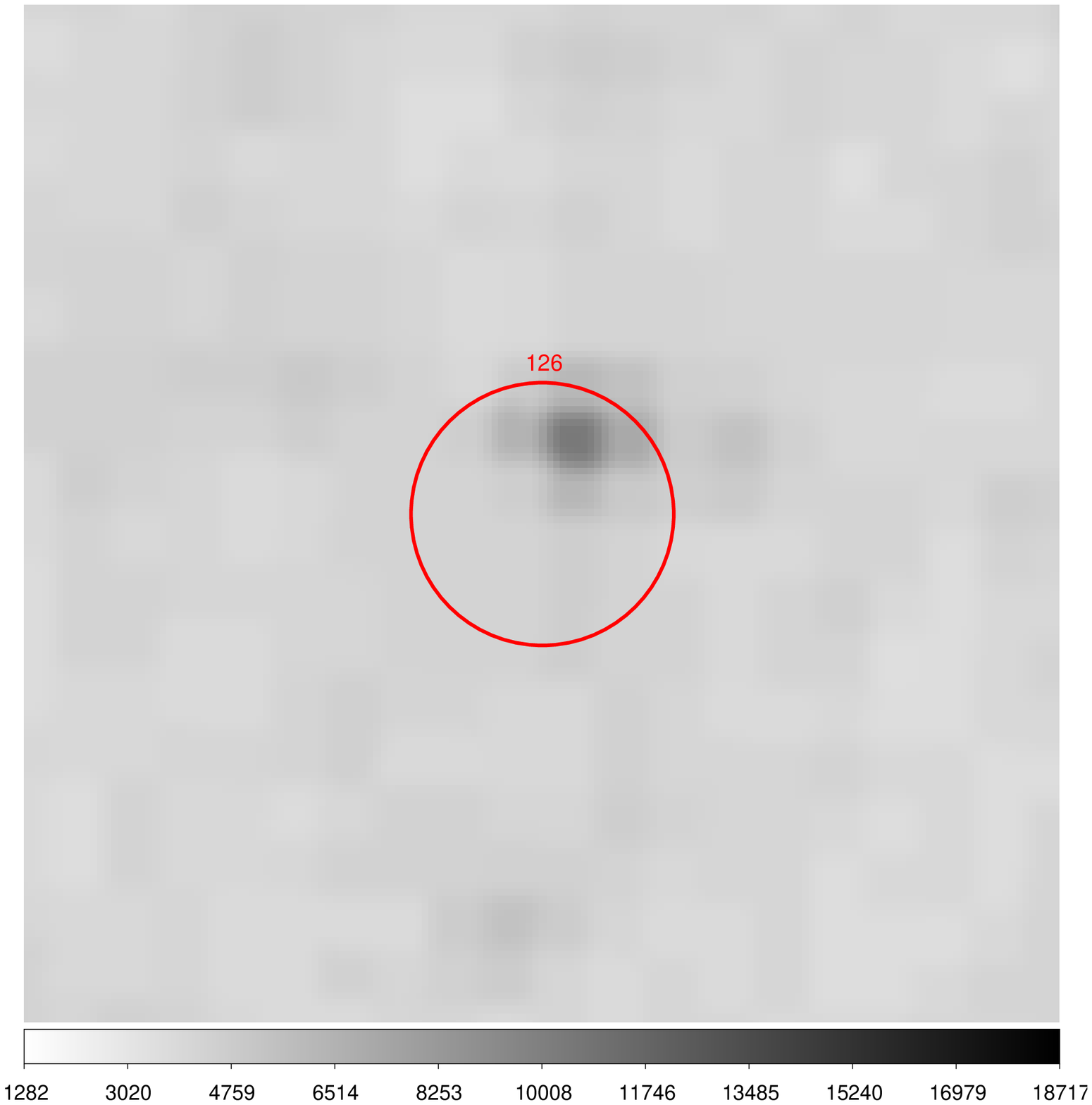}}  \\
  \subfloat[Src-No.\,127  ]{\includegraphics[clip, trim={0.0cm 2.cm 0.cm 0.0cm},width=0.26\textwidth]{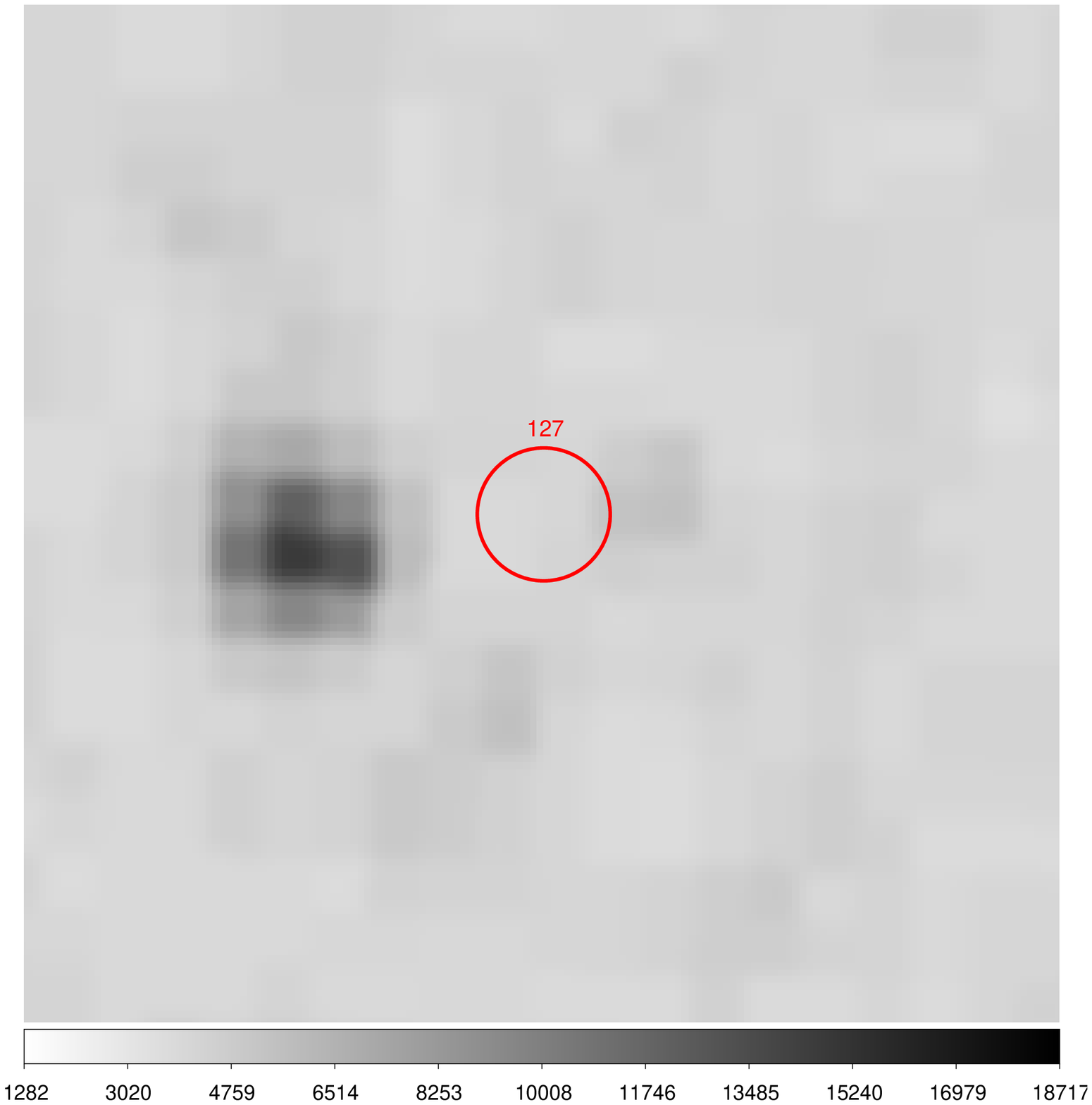}}  
  
\end{figure}

%
\begin{figure}
\caption{Lomb-Scargle periodogram of the bright sources, which are candidates for X-ray sources in Sculptor dSph. The dashed line in the plots shows >3$\sigma$ significance  corresponding to a false alarm probability of 0.001.\label{lomb}}
\centering
 \subfloat[Src-No.\,23 ]{\includegraphics[clip, trim={0.0cm 0.cm 0.cm 0.0cm},width=0.30\textwidth]{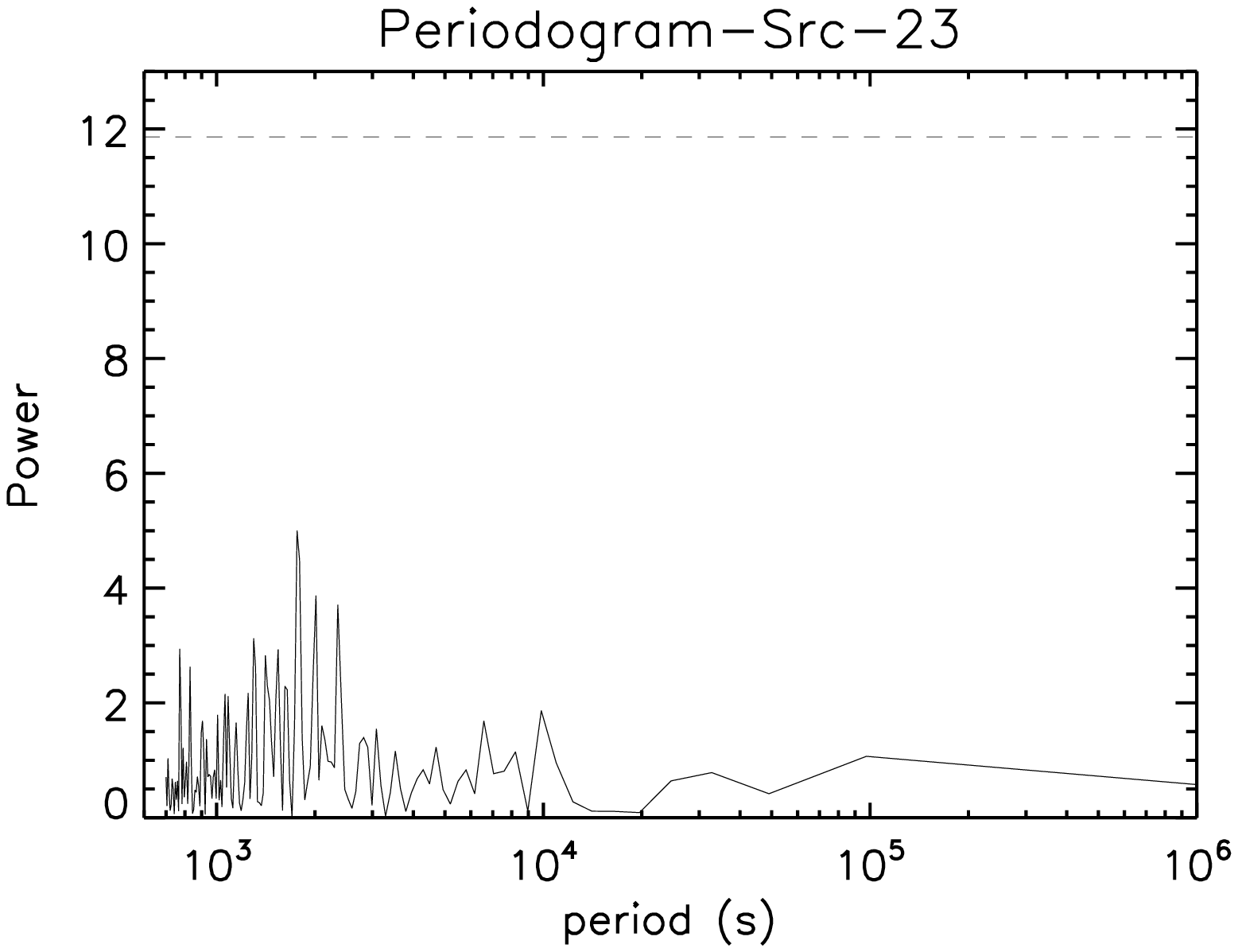}}  
  \subfloat[Src-No.\,27  ]{\includegraphics[clip, trim={0.0cm 0.cm 0.cm 0.0cm},width=0.30\textwidth]{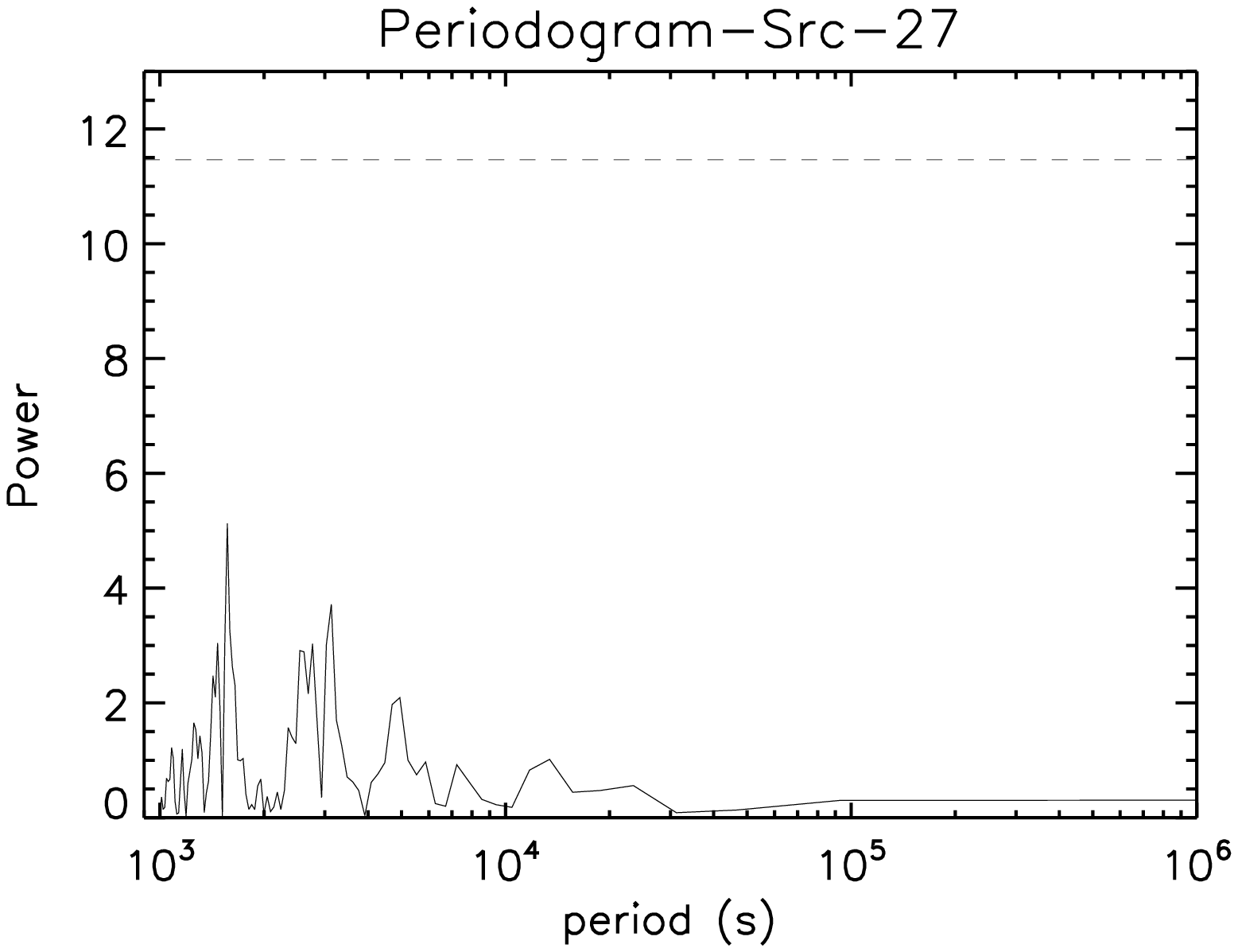}}
  \subfloat[Src-No.\,52  ]{\includegraphics[clip, trim={0.0cm 0.cm 0.cm 0.0cm},width=0.30\textwidth]{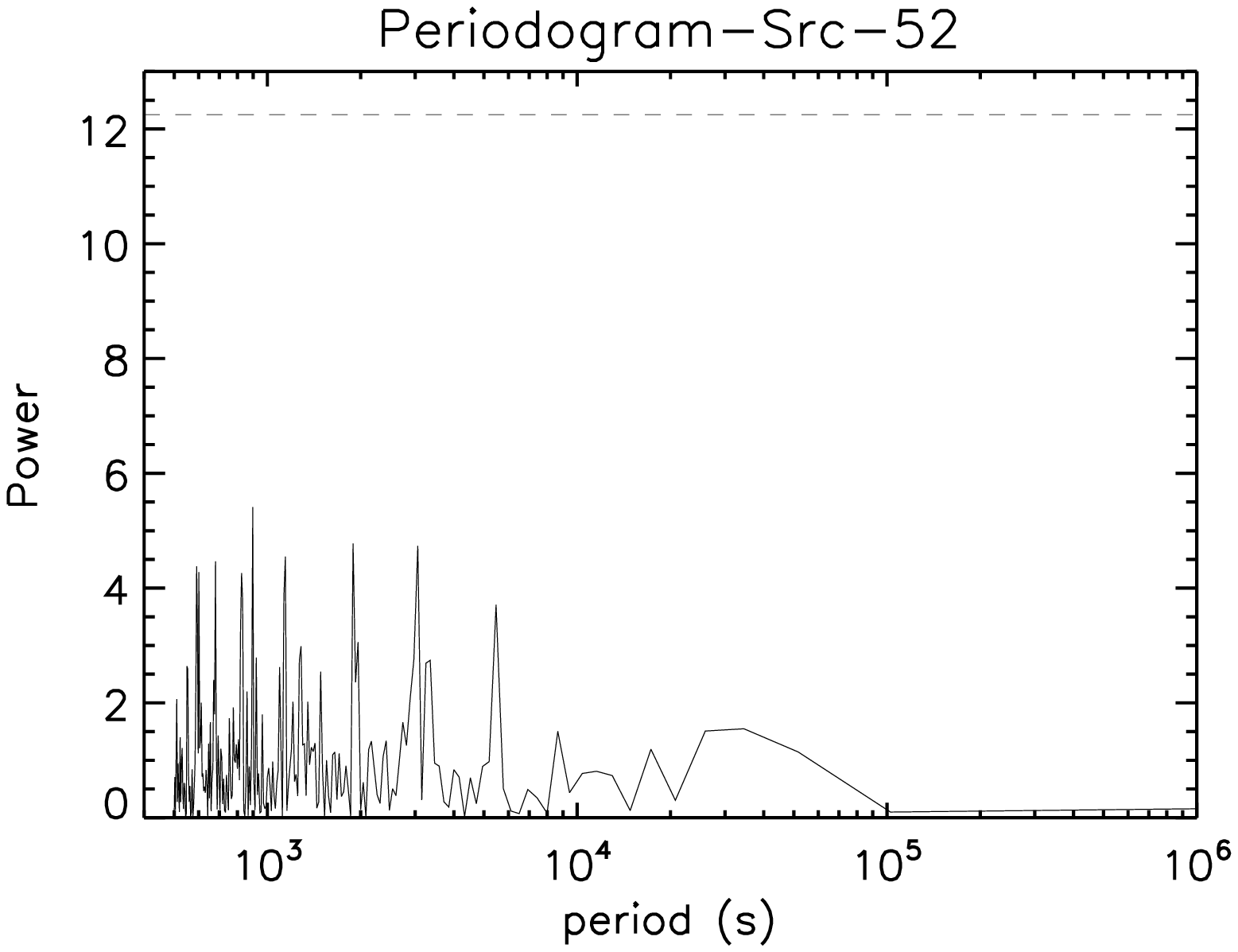}} \\ 
   \subfloat[Src-No.\,69  ]{\includegraphics[clip, trim={0.0cm 0.cm 0.cm 0.0cm},width=0.30\textwidth]{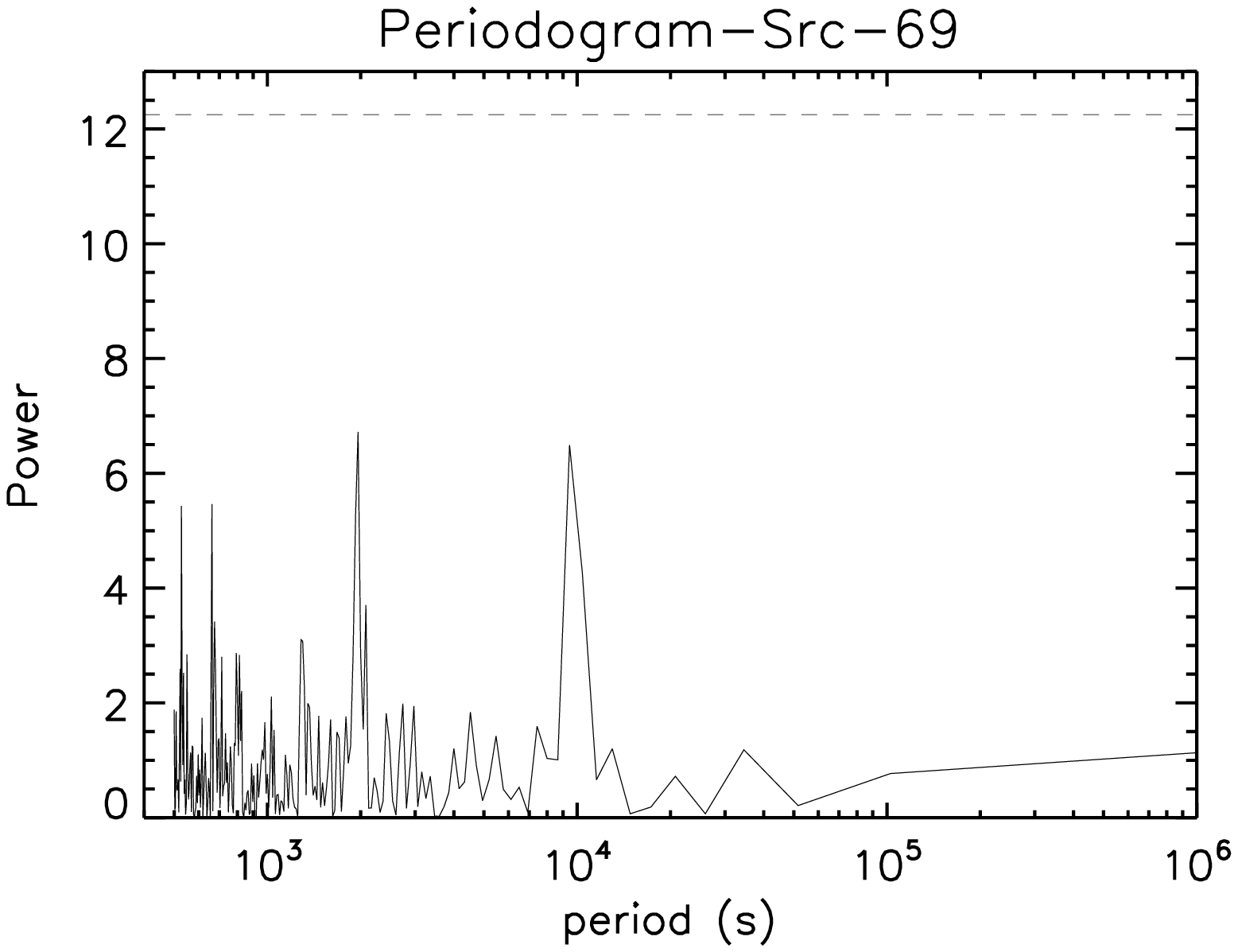}}  
  \subfloat[Src-No.\,74 ]{\includegraphics[clip, trim={0.0cm 0.cm 0.cm 0.0cm},width=0.30\textwidth]{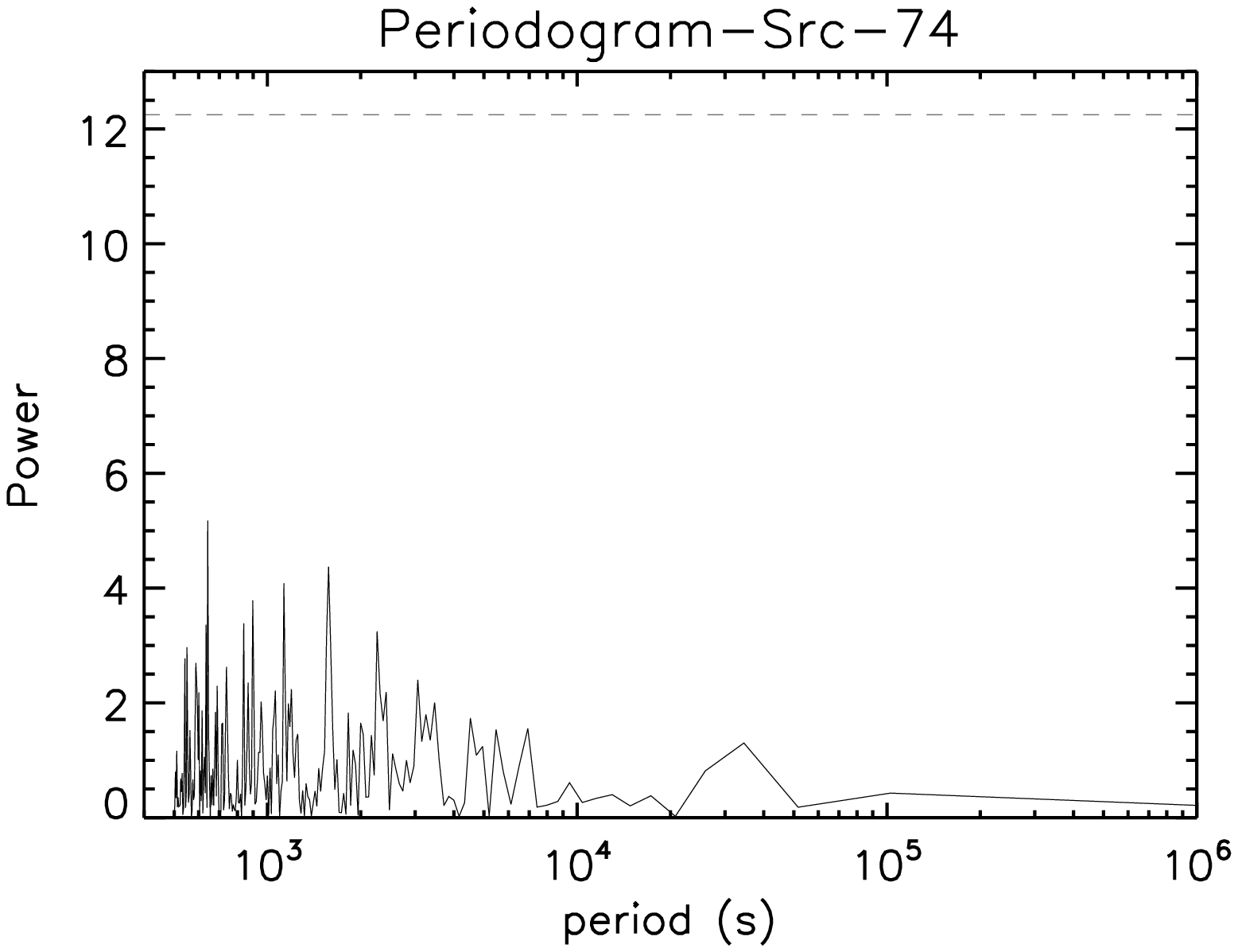}}
  \subfloat[Src-No.\,109 ]{\includegraphics[clip, trim={0.0cm 0.cm 0.cm 0.0cm},width=0.30\textwidth]{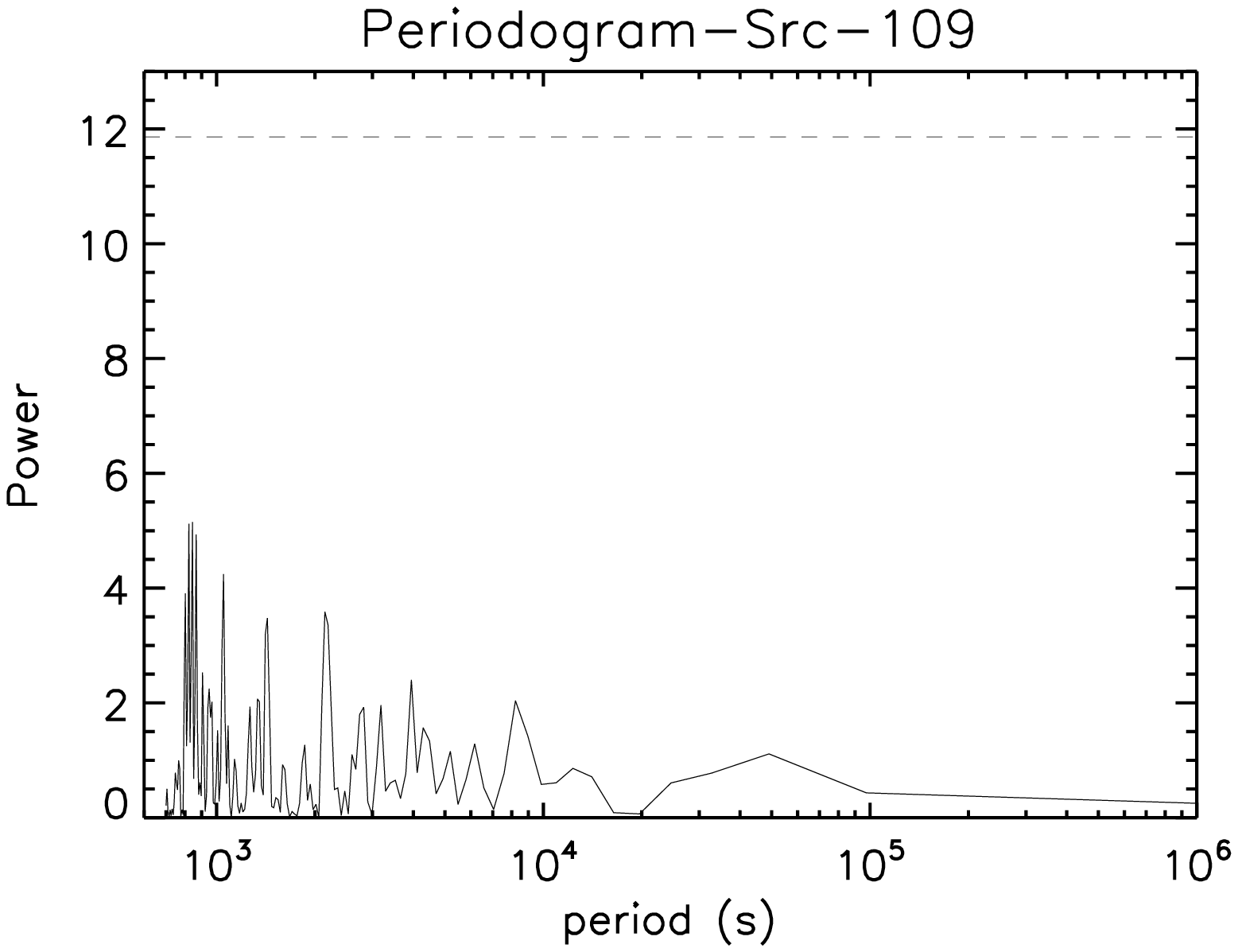}} \\ 
  \subfloat[Src-No.\,127 ]{\includegraphics[clip, trim={0.0cm 0.cm 0.cm 0.0cm},width=0.30\textwidth]{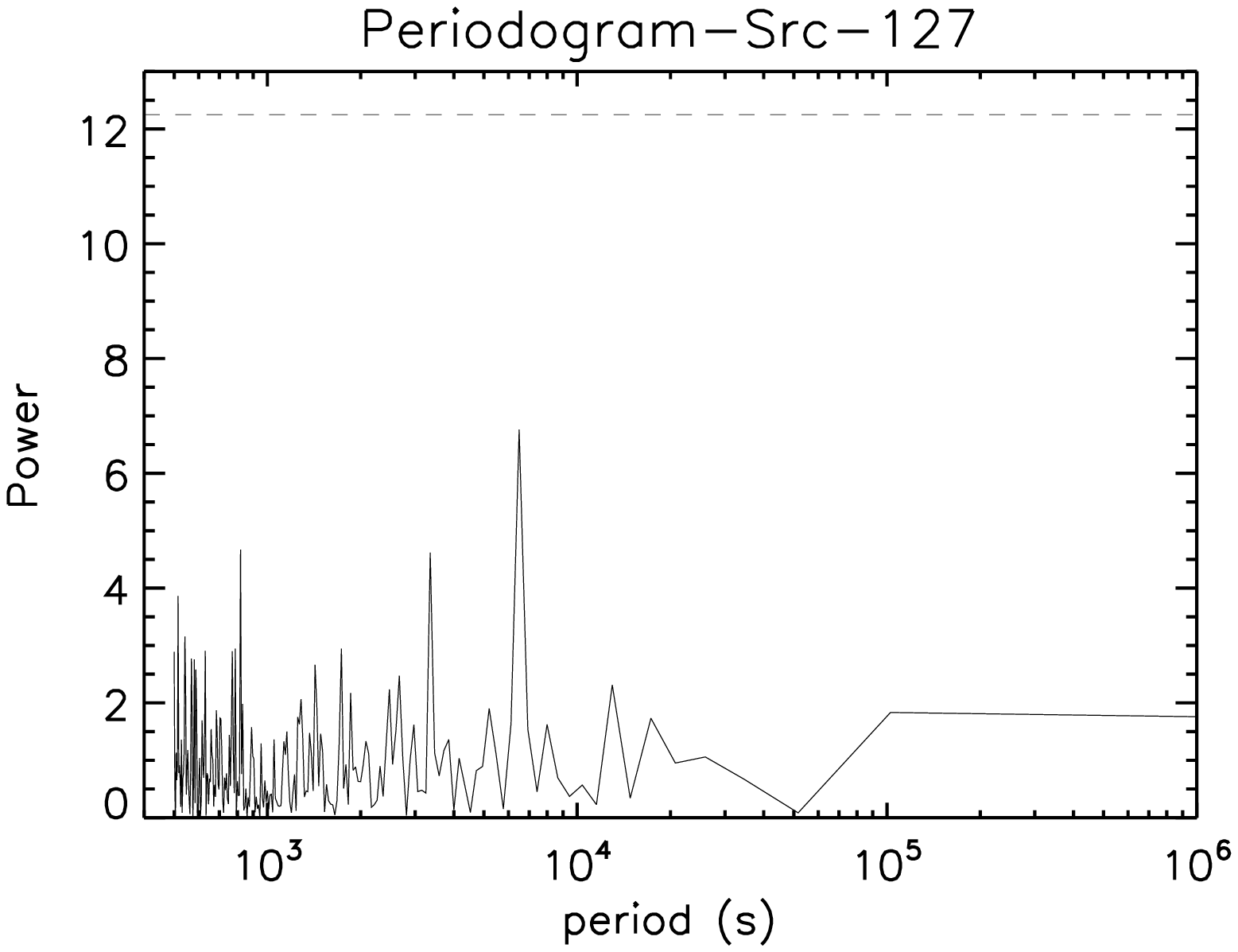}}
  \end{figure}
  
\bsp	
\label{lastpage}
\end{document}